\documentclass[traditabstract, twocolumns,longauth,colorlinks=true,linkcolor=blue,citecolor=blue, urlcolor=blue]{aa}
\usepackage{graphicx}
\usepackage{orcidlink}
\usepackage{txfonts}
\usepackage{amsmath}
\usepackage{caption,subcaption}
\usepackage{placeins}
\usepackage{mathtools}

\usepackage{mathrsfs}
\usepackage{natbib}
\usepackage{xspace}
\usepackage{xcolor}
\usepackage{scalerel}
\usepackage[normalem]{ulem}
\usepackage{enumitem} 
\usepackage{multirow}
\usepackage{threeparttable, tablefootnote}
\usepackage{longtable}
\usepackage{aas_macros}
\usepackage{makecell}

\newcommand{\HEonze}{HE\,1104$-$1805\xspace}
\newcommand{\pycs}{{\tt PyCS}\xspace}
\newcommand{\hc}{\ensuremath{H_{\rm 0}}\xspace }
\newcommand{\lcdm}{\ensuremath{\mathrm{\Lambda CDM}}\xspace}
\newcommand{\kext}{\ensuremath{\kappa_{\rm ext}}\xspace}
\newcommand{\ddt}{\ensuremath{D_{\Delta t}}}
\newcommand{\dd}{\ensuremath{D_{\rm d}}}
\newcommand{\fig}{Fig.}

\newcommand{\kmsmpc}{\ensuremath{\mathrm{km} \ \mathrm{s}^{-1}  \mathrm{Mpc}^{-1}}}

\newcommand*\diff{\mathop{}\!\mathrm{d}}

\newcommand{\gext}{\ensuremath{\gamma_{\rm ext}} \xspace}
\newcommand{\bani}{\ensuremath{\beta_{\rm ani}}}

\newcommand{\lint}{\ensuremath{\lambda_{\rm int}}\xspace}
\newcommand{\slos}{\ensuremath{\sigma^{\rm LOS}}\xspace}

\newcommand{\kms}{\ensuremath{\mathrm{km} \ \mathrm{s}^{-1}}}

\newcommand{\toppanel}{Top panel: }
\newcommand{\bottompanel}{Bottom panel:}
\newcommand{\rightpanel}{Right panel:}
\newcommand{\leftpanel}{Left panel: }

\newcommand{\dtab}{\ensuremath{\Delta t_{\rm AB}}}

\newcommand{\thetae}{\ensuremath{\theta_{\rm E}}\xspace }
\newcommand{\arcs}{\ensuremath{^{\prime \prime}}}

\def\apjl{ApJ}         
\def\apjs{ApJS}         
\def\ao{Appl.~Opt.}     
         
\def\aap{A\&A}

\begin{document} 

\title{TDCOSMO}
\subtitle{XXIII. Measurement of the Hubble constant from the doubly lensed quasar \HEonze}

\titlerunning{Measurement of the Hubble constant from the doubly lensed quasar \HEonze}

   \author{
   Eric~Paic \orcidlink{0000-0002-4306-7366}\inst{\ref{utokyo},\ref{epfl}}\fnmsep\thanks{Email: \href{mailto:ericpaic@g.ecc.u-tokyo.ac.jp}{ericpaic@g.ecc.u-tokyo.ac.jp}}, 
    Fr\'ed\'eric~Courbin\orcidlink{0000-0003-0758-6510}\inst{\ref{iccub},\ref{icrea}}, 
    Christopher~D.~Fassnacht\orcidlink{0000-0002-4030-5461}\inst{\ref{ucdavis}}, 
   Aymeric~Galan\orcidlink{0000-0003-2547-9815}\inst{\ref{tum},\ref{mpi}},     
   Martin~Millon\orcidlink{0000-0001-7051-497X}\inst{\ref{eth}}, 
    Dominique~Sluse\orcidlink{0000-0001-6116-2095}\inst{\ref{uliege}},
    Devon~M.~Williams\orcidlink{0000-0002-8386-0051}\inst{\ref{ucla}},    
   Simon~Birrer\orcidlink{0000-0003-3195-5507}\inst{\ref{stonybrook}},
   Elizabeth~J.~Buckley-Geer\orcidlink{0000-0002-3304-0733}\inst{\ref{fnal},\ref{uchicago}},
   Michele~Cappellari\orcidlink{0000-0002-1283-8420}\inst{\ref{oxford}}, 
   Fr\'ed\'eric~Dux\orcidlink{0000-0003-3358-4834}\inst{\ref{epfl},\ref{eso}},
   Xiang-Yu~Huang\orcidlink{0000-0001-7113-0599}\inst{\ref{stonybrook}},
   Shawn~Knabel\orcidlink{0000-0001-5110-6241}\inst{\ref{ucla}},
   Cameron~Lemon\orcidlink{0000-0003-2456-9317}\inst{\ref{okc}},
   Anowar~J.~Shajib\orcidlink{0000-0002-5558-888X}\inst{\ref{uchicago},\ref{kicp},\ref{cassa}},
   Sherry H.~Suyu\orcidlink{0000-0001-5568-6052}\inst{\ref{tum},\ref{mpi}},
   Tommaso~Treu\orcidlink{0000-0002-8460-0390}\inst{\ref{ucla}},
   Kenneth~C.~Wong\orcidlink{0000-0002-8459-7793}\inst{\ref{utokyo}},
   Lise Christensen\orcidlink{0000-0001-8415-7547}\inst{\ref{dawn}},
   Veronica Motta\orcidlink{0000-0003-4446-7465}\inst{\ref{valparaiso}},
   Alessandro~Sonnenfeld\orcidlink{0000-0002-6061-5977}\inst{\ref{shanghai1},\ref{shanghai2},\ref{shanghai3}}
   }
   \institute{
   Research Center for the Early Universe, Graduate School of Science, The University of Tokyo, 7-3-1 Hongo, Bunkyo-ku, Tokyo 113-0033, Japan
    \label{utokyo}\goodbreak
    \and
    Institute of Physics, Laboratory of Astrophysics, Ecole Polytechnique F\'ed\'erale de Lausanne (EPFL), Observatoire de Sauverny, 1290 Versoix, Switzerland \label{epfl}\goodbreak
    \and
    Institut de Ci\`{e}ncies del Cosmos (ICCUB), 
Universitat de Barcelona (IEEC-UB), 
Mart\'{i} i Franqu\`{e}s 1, 08028 Barcelona, Spain
\label{iccub}\goodbreak
\and
Instituci\'o Catalana de Recerca i Estudis Avan\c{c}ats (ICREA), 
Passeig de Llu\'{\i}s Companys 23, 08010 Barcelona, Spain
\label{icrea}\goodbreak
\and
    Department of Physics and Astronomy, UC Davis, 1 Shields Ave., Davis, CA 95616, USA
    \label{ucdavis}\goodbreak
    \and
    Technical University of Munich, TUM School of Natural Sciences, Physics Department,  James-Franck-Stra{\ss}e 1, 85748 Garching, Germany 
    \label{tum}\goodbreak
    \and
    Max-Planck-Institut f{\"u}r Astrophysik, Karl-Schwarzschild Stra{\ss}e 1, 85748 Garching, Germany
    \label{mpi}\goodbreak
    \and
    Institute for Particle Physics and Astrophysics, ETH Zurich, Wolfgang-Pauli-Strasse 27, CH-8093 Zurich, Switzerland 
        \label{eth}\goodbreak
    \and
    STAR Institute, Li\`ege Universit\'e, Quartier Agora - All\'ee du six Ao\^ut, 19c B-4000 Li\`ege, Belgium
    \label{uliege}\goodbreak
    \and 
    Department of Physics and Astronomy, University of California, Los Angeles, CA 90095, USA
    \label{ucla}\goodbreak
    \and
    Department of Physics and Astronomy, Stony Brook University, Stony Brook, NY 11794, USA \label{stonybrook}\goodbreak
    \and
    Fermi National Accelerator Laboratory, PO Box 500, Batavia, IL 60510, USA 
    \label{fnal}\goodbreak
    \and
    Department  of  Astronomy  \&  Astrophysics,  University  of Chicago, Chicago, IL 60637, USA; \label{uchicago}\goodbreak
    \and
    Sub-Department of Astrophysics, Department of Physics, University of Oxford, Denys Wilkinson Building, Keble Road, Oxford, OX1 3RH, UK
    \label{oxford}\goodbreak
    \and
    European Southern Observatory, Alonso de Córdova 3107, Vitacura, Santiago, Chile
    \label{eso}\goodbreak
    \and
    Oskar Klein Centre, Department of Physics, Stockholm University, SE-106 91, Stockholm, Sweden
    \label{okc}\goodbreak
   \and
    Kavli Institute for Cosmological Physics, University of Chicago, Chicago, IL 60637, USA \label{kicp}\goodbreak
    \and
    Center for Astronomy, Space Science and Astrophysics, Independent University, Bangladesh, Dhaka 1229, Bangladesh \label{cassa}\goodbreak
    \and 
    Cosmic Dawn Center (DAWN), Niels Bohr Institute, University of Copenhagen, Jagtvej 128, DK-2200 Copenhagen N, Denmark
    \label{dawn} \goodbreak
    \and 
    Instituto de F\`{\i}sica y Astronom\'{\i}a, Universidad de Valpara\'{\i}so, Av. Gran Breta\~na 1111 Playa Ancha, Valpara\'{\i}so, Chile
    \label{valparaiso} \goodbreak
    \and 
    Department of Astronomy, School of Physics and Astronomy, Shanghai Jiao Tong University, Shanghai 200240, China \label{shanghai1} \goodbreak
    \and 
    Shanghai Key Laboratory for Particle Physics and Cosmology, Shanghai Jiao Tong University, Shanghai 200240, China
    \label{shanghai2} \goodbreak
    \and 
    Key Laboratory for Particle Physics, Astrophysics and Cosmology, Ministry of     Education, Shanghai Jiao Tong University, Shanghai 200240, China
    \label{shanghai3} 
    }

\abstract{
Time-delay cosmography leverages strongly lensed quasars to measure the Universe's current expansion rate, \hc, independently from other methods. The latest TDCOSMO milestone measurement primarily used quadruply lensed quasars for their mass profile constraints. However, doubly lensed quasars, being more abundant and offering precise time delays, could expand the sample by a factor of 5, significantly advancing towards a 1\% precision measurement of \hc.
  
We present the first TDCOSMO analysis of a doubly imaged source, \HEonze, including the measurement of the four necessary ingredients.
First, by combining 17 years of data from the SMARTS, Euler, and WFI telescopes, we
measured a time delay of 176.3$^{+11.4}_{-10.3}$ days. Second, using MUSE data, we extracted stellar velocity dispersion measurements in three radial bins with 5\% to 13\% precision. Third, employing F160W HST imaging for lens modelling and marginalising over various modelling choices, we measured the Fermat potential difference between the images. 
Fourth, using wide-field imaging, we measured the convergence added by objects not included in the lens modelling. 
By combining these four ingredients, we measured the time delay distance and the angular diameter distance to the deflector, favouring a power-law mass model over a baryonic and dark matter composite model. The measurement was performed blindly to prevent experimenter bias and resulted in a Hubble constant of $\hc = 64.2^{+5.8}_{-5.0} $ $\times$ \lint \kmsmpc, where \lint is the internal mass sheet degeneracy parameter. This is in agreement with the TDCOSMO-2025 milestone and its precision for $\lint=1$ is comparable to that obtained with the best-observed quadruply lensed quasars (4-6\%).

This work is a stepping stone towards a precise measurement of \hc using a large sample of doubly lensed quasars, supplementing the current sample. The next TDCOSMO milestone paper will include this system in its hierarchical analysis, constraining $\lint$ and \hc\ jointly with multiple lenses. 
}

\authorrunning{Paic et al.}

 \maketitle
\section{Introduction}
\label{sec:introduction}

In recent years, the  $\Lambda$ cold dark matter (\lcdm) cosmological paradigm has been challenged by observations, as tensions on several cosmological parameters now reach a statistically significant level. The most significant of these tensions is about the Hubble constant \hc, which gives the current expansion rate of the Universe. While the most recent \hc measurements using the cosmic microwave background (CMB) obtain \hc = $67.4\pm0.5$ \kmsmpc \ \citep{Planck2020}, local measurements find higher values, \citep[e.g.][]{Verde2019,Cosmoverse25}. In particular, the distance ladder using parallaxes, water megamasers, Cepheids, and type-Ia supernovae yields \hc = $73.17\pm0.86$ \kmsmpc\ \citep{breuval24}, which is in a $5\sigma$ tension with the CMB value. Given this discrepancy, it is essential to develop alternative cosmological probes that are fully independent of those listed above and that are competitive in terms of precision and accuracy.

Strong gravitational lensing occurs when a massive and compact object (the lens) intersects the line-of-sight (LOS) to a more distant galaxy, forming multiple images of the same background source. The lensed images appear at positions on the sky that minimise the photons' travel time to the observer, following Fermat's principle. These optical paths differ in length, resulting in a time delay between the arrival times of the photon in the different multiple images. These time delays are proportional to the so-called time delay distance \ddt\ \citep[e.g.][]{refsdal64,suyu14}, a quantity that is inversely proportional to \hc. The method, first proposed by \citet{refsdal64}, requires the time-delay measurement and modelling of the mass along the LOS up to the source redshift, including the main deflector while breaking the inherent mass-sheet degeneracy \citep[MSD, e.g.][]{Falco1985,Gorenstein1988,Kochanek2002, Schneider2013,schneider14, Blum2021}. 

The TDCOSMO collaboration \citep[e.g.][]{millon20c,birrer20,shajib23} is working towards a 1\% precision measurement of \hc using time-delay cosmography (TDC) with strongly lensed quasars, one of the few cosmological probes that are truly independent of the local distance ladder. 
The initial set of \hc-measurements by our TDCOSMO collaboration \citep[e.g.][]{suyu10,suyu17, bonvin17,millon20c, Wong2020} used empirically motivated parametric models, which is an 'assertive' \citep{TSM22} model assumption consistent with several independent observations from stellar dynamics \citep{cappellari16} or galaxy simulations \citep{navarro97}, and also from joint lensing-dynamics studies \citep{shajib21}. However, this assertive model assumption implicitly breaks the MSD, and thus risks underestimating the uncertainty beyond the sole constraining power of the data.  Therefore, TDCOSMO-IV \citep{birrer20} incorporates one additional degree of freedom in the mass model that is maximally degenerate with the MSD and constrains this additional degree of freedom using stellar kinematics, which breaks the MSD.
As expected, the precision of the measurement decreases, but the value remains compatible with previous ones. Recently, spatially resolved dynamics of the lensing galaxy yielded from integral field unit (IFU) data have been shown to significantly improve the constraints on the mass profile and the \hc-measurement on individual lenses \citep{shajib18, Yildirim2020, shajib23}. 
 
The latest milestone measurement of the collaboration \citep[][hereafter TDCOSMO25]{tdcosmo25} used new IFU data from the JWST, Keck, and VLT telescopes with the improved methodology of \citet{knabel25}.  The hierarchical inference, jointly modelling all TDCOSMO lenses together with a subset of the Sloan Lenses ACS and Strong Lenses in the Legacy Survey samples, in combination with information on $\Omega_{\rm m}$ from independent datasets such as the Pantheon Supernovae catalog \citep{brout22} or DESI's baryonic acoustic oscillation measurement \citep{desi25} on the TDCOSMO lenses, yields respectively $\hc = 74.3^{+3.1}_{-3.7}\ \kmsmpc$ and $\hc = 74.8^{+3.5}_{-3.4}\ \kmsmpc$.  This milestone marks significant improvements in both accuracy and precision. One of the main ways to further improve precision is to increase the number of time-delay lenses in the sample \citep{B+T21}.

As they are more magnified, quadruply lensed quasars are easier to find serendipitously than doubles. Hence, lensed quasar datasets were initially dominated by them since the image multiplicity and the three independent delays also give more constraints on the mass modelling of the main deflector. However, these systems are rare in the sky, and one of the current limiting factors of the measurements is the sample size. As wide-field surveys emerged with deeper imaging, the discovery selection bias was attenuated \citep{lemon20,lemon23, dux24, lemon24} and doubles now represent 3/4 of known lensed quasars. They are generally easier to monitor from the ground because of their wider separation, and therefore have the potential to greatly reduce statistical noise. The extension of the TDCOSMO sample to doubles will accelerate progress towards the ultimate goal of 1\% precision \citep{TSM22}.

Furthermore, quads generally require more exquisite alignment of deflector and source, and higher ellipticity/shear than doubles. Therefore, restricting the TDCOSMO sample to quads may expose the \hc \ measurement to unknown selection biases \citep{ sonnenfeld23}. Analysing a sample of doubly lensed quasars offers a way to improve statistical precision and unveil and mitigate any potential biases that may be present. 
 
The present paper is the first TDCOSMO paper to analyse a doubly imaged quasar \footnote{In the system SDSS\,1206$+$4332, analysed by \citet{birrer19}, the quasar is doubly imaged while its host galaxy was imaged four times. Hence, it gives more constraints on the mass model than a standard double, such as \HEonze}, namely \HEonze discovered by \citet{wisotzki93} with a source at redshift $z_{\rm s}=2.32$ and deflector at redshift $z_{\rm d} = 0.729$ 
\citep{wisotzki93, courbin98, lopez99, lidman00}. 
Following the collaboration protocol, the measurement was performed blindly to prevent experimenter bias. The mean value of the posterior distribution function of the Hubble Constant was kept hidden from the investigators and unblinded on July 8th 2025, during a collaboration teleconference. The paper was mostly written before unblinding, with placeholder figures and tables, and completed after unblinding. The unblinded results were published without modification.

The paper outline is as follows. Sect.~\ref{sec:TDC} introduces the basics of TDC.
The time-delay measurement is presented in Sect.~\ref{sec:TD} before turning to the spatially resolved kinematics of the lens with the ESO-VLT MUSE instrument in Sect.~\ref{sec:MUSE}. Sect.~\ref{sec:los} describes the mass contribution of galaxies along the LOS, and Sect.~\ref{sec:models} shows the different models we developed to measure the time delay distance to \HEonze. Sect.~\ref{sec:inference} gives our cosmological inference and \hc\ value after unblinding of the whole analysis chain. Finally, Sect.~\ref{sec:discussion} provides the limitations of our work, and we conclude in Sect.~\ref{sec:conclusion}.

\section{Time-delay cosmography formalism}
\label{sec:TDC}

The determination of \hc using the time delay observable in strong lens systems was first suggested by \citet{refsdal64} and further developed by \citet{schneider92}. In this section, we review the main points of this formalism, highlighting how the redshifts of the source ($z_{\rm s}$) and the lens ($z_{\rm d}$), the time delay between each image (\dtab), the resolved kinematics, the LOS, the mass model of the lens and the position of the source are used to measure \hc. \footnote{We refer the interested reader to the extensive reviews of \citet{ISSI-Birrer24} and \citet{Treu2023} for more details.}

\subsection{Lensing formalism}

The time delay $\dtab$ between two images $A$ and $B$ of a given strong lens system is given byinfer
\begin{align}
    \dtab &= \frac{\ddt}{c} \left(\tau\left(\vec{\theta_{\rm A}}, \vec{\beta}\right)- \tau\left(\vec{\theta_{\rm B}}, \vec{\beta}\right)\right),\\
    \text{with } \ddt &\equiv (1+z_{\rm d})\frac{\dd D_{\rm s}}{D_{\rm d s}}, \\
    \text{and } \tau\left(\vec{\theta}, \vec{\beta}\right) &= \frac{1}{2}\left(\vec{\theta} -\vec{\beta} \right)^2 - \phi\left(\vec{\theta}\right),
    \label{eq:dt}
    \end{align}

where $\vec{\theta}$ is the apparent angular position of an image and $\vec{\beta}$ the position of the source, \ddt\ is the so-called 'time-delay distance' defined as a function of the lens redshift $z_{\rm d}$, and the angular distances \dd, $D_{\rm s}$ and $D_{\rm ds}$, respectively, between the observer and the lens, the observer and the source, and the lens and the source. Finally, the Fermat potential $\tau(\vec{\theta}, \vec{\beta})$ depends on the lens potential $\phi(\vec{\theta})$, which can be determined by modelling the mass distribution of the lens. 

We can generalise Eq.~\ref{eq:dt} in the case where multiple lenses lie at different redshifts, creating $P$ different lens planes. The time delay then becomes \citep[e.g. ][]{schneider92}

\begin{align}
    \dtab = \sum ^{\rm P}_{ i=1} \frac{D_{\Delta t,i,i+1}}{c} \left[\frac{(\vec{\theta}_{\rm A, \textit{i}}-\vec{\theta}_{\rm A,\textit{i+1}})^2}{2} -  \frac{\vec{(\theta}_{\rm B,\textit{i}}-\vec{\theta}_{\rm B,\textit{i+1}})^2}{2} \right. \nonumber \\
    \left. - \zeta_{i,i+1} \left(\phi_{i}(\vec{\theta}_{\rm A, \textit{i}}) + \phi_{i}(\vec{\theta}_{\rm B, \textit{i}})\right) \vphantom{\frac{D_{\Delta t,i,i+1}}{c}}\right],
    \label{eq:multiplane}
\end{align}

\begin{equation}
    D_{\Delta t,i,i+1} \equiv {(1+z_{i})}\frac{D_{i} D_{i+1}}{D_{ ii+1}}, \ \ \ \ \ 
\end{equation}

\begin{equation}
    \zeta_{i,i+1} \equiv \frac{D_{i,i+1}D_{\rm s}}{D_{i+1} D_{\rm is}} ,
\end{equation}

 By defining $D_{\Delta t,{\rm eff}} \equiv D_{\Delta t,l,s}$ as the time-delay distance between the main lens plane and the source plane, the time delay can be rewritten
\begin{equation}
    \dtab = \frac{\ddt^{\rm eff}}{c} \Delta \phi_{\rm{AB}}^{\rm{eff}} .
\end{equation}

With the effective Fermat potential 
\begin{equation}
    \Delta \phi^{\rm eff}(\theta) = \sum^{\rm P}_{i=1} \frac{1+z_{i}}{1+z_{\rm d}}\frac{D_{i} D_{i+1}D_{\rm ds}}{D_{\rm d} D_{\rm s} D_{i,i+1}} \left[\frac{(\theta_{i} - \theta_{i+1})^2}{2} - \zeta_{i,i+1} \psi_{i}(\theta_{i}) \right].
    \label{eq:fermatpotmultiplane}
\end{equation} 

As pointed out by \citet{Falco1985}, the observables of a strong lensing system, such as the image position and the flux ratios, are invariant with respect to the mass-sheet transform (MST). This transformation of the convergence $\kappa(\theta)$ and the source plane coordinates $\vec{\beta}$ can be written as 
\begin{align}
    \label{eq:MSD}
    \kappa_{\lambda}\left(\theta\right) &= \lambda \kappa\left(\theta\right) + (1 -\lambda) \ , \\
    \vec{\beta_{\lambda}} &= \lambda\vec{\beta}.
\end{align}
This degeneracy between the mass of the lens and the size of the source is called the mass-sheet degeneracy (MSD) and implies a degeneracy between \hc and the time delay. 

Although a pure mass sheet extending up to infinity in the lens plane is unphysical, the MSD is the manifestation of two physical effects: 1) the lensing effect of all the galaxies along the LOS other than the main lens. We refer to this effect as the external MSD, and it can be treated as equivalent to a mass sheet with a surface mass density \kext in units of the critical density for lensing; 2) the internal mass sheet degeneracy, which results from a small mismatch between the true mass profile of the lens and its model. 

The total internal and external MSD parameters defined in Eq.~\ref{eq:MSD} can therefore be decomposed into two terms 
\begin{equation}
    \lambda= (1 - \kext) \lint, 
    \label{eq:lambda}
\end{equation}
where \lint is the internal MSD parameter. 

The external part of the MSD can be statistically derived if \kext is estimated from the mass of the galaxies along the LOS. This can be done either by counting the galaxies in a large aperture and comparing the density of the LOS to numerical simulations \citep[see e.g.][for detail]{wells23} or with weak lensing \citep{Tihhonova2018, Tihhonova2020}. The time-delay distance \ddt\ can then be corrected $\ddt\ = \ddt\mathrm{'}/\left(1-\kext\right)$, where $\ddt\mathrm{'}$ is the time-delay distance from mass modelling without accounting for $\kext$. The detailed computation of \kext is presented in Section \ref{sec:kextmeas}. 

Following previous TDCOSMO single-lens analyses \citep{wong17, rusu20, shajib20, wong24}, we assumed no internal MST (i.e. \lint=1). The next collaboration milestone that will hierarchically model all TDCOSMO lenses will release this parameter.

\subsection{Kinematic modelling \label{sec:kinmod}}

The kinematic observable is the luminosity-weighted line-of-sight projected stellar velocity dispersion, denoted as \slos. It is measured by targeting stellar absorption lines and quantifying their width with high-resolution spectra.

As in TDCOSMO25, we used the spherically aligned Jeans anisotropic modelling (JAM) method. This method builds on the orbital distribution $f(\vec{x}, \vec{v})$ of position and velocity of the stars in 3D motion in the galactic potential $\Phi$. This potential is described by the steady-state collisionless Boltzmann equation \citep[][hereafter, BT, eq.~4-13b)]{binne87} 
\begin{equation}
    \sum^3_{i=1} \left(v_{  i}\frac{\partial f}{\partial x_{  i}}- \frac{\partial \Phi}{\partial x_{  i}}\frac{\partial f}{\partial v_{  i}} \right) = 0.
\end{equation}

We then obtain a single spherical Jeans equation (BT eq.~4-54) 
 \begin{equation}
     \frac{d \left[\rho_*(r)\sigma^2_{r}(r)\right]}{d r} + \frac{2\bani(r)\rho_*(r)\sigma_{  r}^2(r)}{r} = -\rho_* \frac{d \Phi(r)}{d r}, 
     \label{eq:jeans}
 \end{equation}
with $\rho_*(r)$ the stellar density distribution and \bani \  the orbital anisotropy, defined as a function of the ratio of tangential over radial velocity dispersion components, $\sigma_{\theta}$ and $\sigma_{r}$ 
\begin{equation}
    \bani \equiv 1 - \frac{\sigma_{\theta}^2}{\sigma_{r}^2}.
\end{equation}
Given $\slos$, the values of $\sigma_{ r}$ and $\sigma_{\theta}$ are degenerate, leaving \bani \ nearly unconstrained. 

This introduces a degeneracy between $\slos$ and the 3D mass profile known as the mass-anisotropy degeneracy (MAD) \citep{binney82, merritt85}. 
As demonstrated in Appendix \ref{app:slowrotator}, the lens galaxy of \HEonze is a slow rotator. We hence followed \citet{tdcosmo25}, and use the measurements of \citet{cappellari07} on slow rotating galaxies to apply the uniform prior $0.87\leq \frac{\sigma_{\theta}^2}{\sigma_{r}^2}\leq 1.12$.

A solution of Eq.~\ref{eq:jeans} is given by 

\begin{equation}
    \sigma^2_{ r}(r) = \frac{G}{\rho_*(r)} \int^{\infty}_{ r} \frac{M(s)\rho_*(s)}{s^2} \mathcal{J}_{ \beta}(r,s) ds,
\end{equation}
with $M(r)$ the mass enclosed within the radius $r$ and 

\begin{equation}
    \mathcal{J}_{ \beta}(r,s) = \exp\left[\int^{ s}_{ r} 2\bani (r') \frac{dr'}{r'}\right]. 
\end{equation}

We then get the modelled velocity dispersion along the LOS, $\slos_{\rm model}$ with (BT, eq.~4-60) 
\begin{equation}
    \left(\slos_{\rm model}\right)^2 =\frac{2}{\Sigma_*(R)} \int_{ R}^{\infty}  \left[1-\bani(r)\frac{R^2}{r^2}\right]\frac{\rho_*(r)\sigma_{ r}^2(r)}{\sqrt{r^2-R^2}}r dr,
    \label{eq:slos}
\end{equation}
with $R$ the projected radius and $\Sigma_*$ the enclosed surface stellar density, which can be constrained from the luminosity profile of the lens $I(R)$ assuming, for instance, a constant mass-to-light ratio 
$\Upsilon$ $\Sigma_*(R)= \Upsilon I(R)$.

To compare this with the observed velocity dispersion along the LOS, \slos, we weigh $\slos_{\rm model}$ with the point spread function (PSF, $\mathcal{P}$) convolved light profile of the lens 
\begin{equation}
    \left(\slos_{\rm sph}\right)^2 = \frac{\int_{\rm A} I(R)\left(\slos_{\rm model}\right)^2 \ast \mathcal{P}  \diff A}{\int_A I(R)\ast \mathcal{P} \diff A} . 
\end{equation}

Following the methodology of TDCOSMO25, we did not account for the inclination angle of the lens galaxy during the JAM modelling. To prevent bias raised by \citet{huang25}, we adopted the \textit{axisymmetric correction factor} $\sigma_{\rm axi}^{\rm ap}/\sigma_{\rm sph}^{\rm ap}$ proposed by these authors. This factor was computed using the shape distribution from \citet{li18} and assuming random inclination. We then bin this factor in the same fashion as the IFU data (see Section \ref{sec:vdispmeas} for more details) to obtain a correction factor for each bin, respectively: $0.992\pm 0.004$, $0.977\pm 0.001$, and $0.993 \pm 0.004$.

The final \slos to be compared with data is then given by 
\begin{equation}
    \slos = \slos_{\rm sph}\cdot \frac{\sigma_{\rm axi}^{\rm ap}}{\sigma_{\rm sph}^{\rm ap}}
    \label{eq:slosfinal}
\end{equation}

The prediction of the observed LOS velocity dispersion from any model, irrespective of the approach, can be decomposed into a cosmology-dependent ratio of angular diameter distances and a cosmology-independent part following \citet{birrer16, birrer19} 
\begin{align}
    \left(\slos\right)^2 &= (1-\kext) \frac{D_{\rm s}}{D_{\rm d s}} \frac{c^2}{J(\xi_{\rm lens},\xi_{\rm light}, \beta_{\rm ani})}, \\
    &= \frac{(1-\kext)}{1+z_{\rm d}} \frac{D_{\rm \Delta t}}{D_{\rm d}} \frac{c^2}{J(\xi_{\rm lens},\xi_{\rm light}, \beta_{\rm ani})},
    \label{eq:sloscosmo}
\end{align}

where the dimensionless quantity $J$ relies on the deflector model parameters ($\xi_{\rm lens}$ and $\xi_{\rm light}$) and $\bani$. Moreover, $J$ considers observational conditions and luminosity-weighting within the dispersion measurement aperture, as previously demonstrated in studies such as \citet{binney82} and \citet{treu04}. 

As the angular diameter distances are sensitive to $\Omega_{\rm m}$ only to the second order, they are primarily sensitive to \hc. Large samples of lenses and combination with probes independent from Time-delay Cosmography are required to constrain other parameters \citep[e.g.][]{linder11,shajib25}.

\begin{figure*}
    \sidecaption
    \minipage{0.44\textwidth}
    \includegraphics[width =\linewidth]{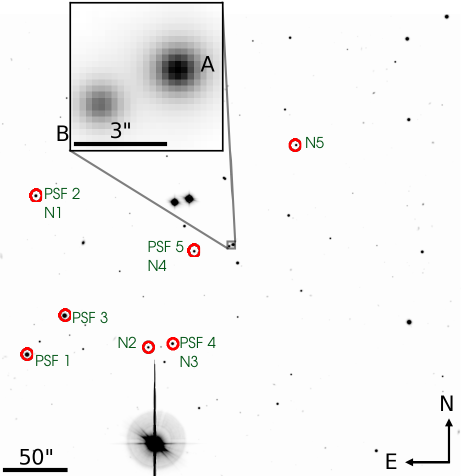}
    \endminipage
    \minipage{0.24\textwidth}
    \includegraphics[width =\linewidth]{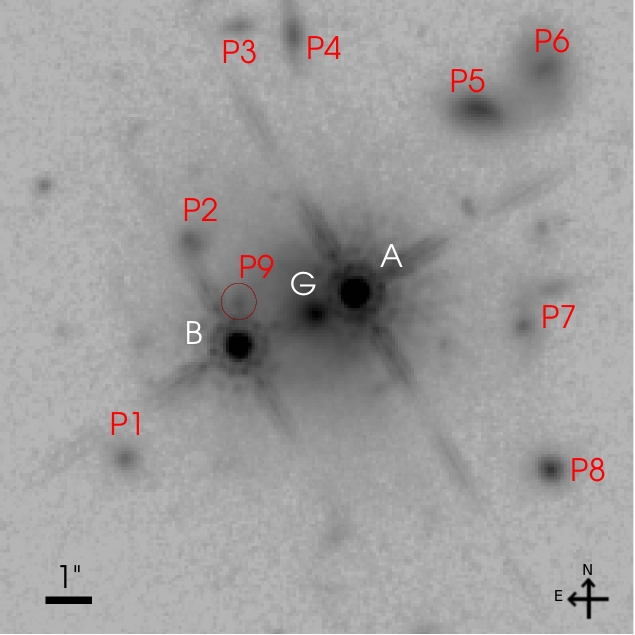}
    \endminipage
    \caption{ \leftpanel Stack of 280 WFI exposures of the field of view used for light curve extraction. Stars nominated PSF1 to PSF5 are used to model the PSF of each exposure, and stars designated as N1 to N5 are used to normalise the flux between each exposure. \rightpanel \ \HEonze imaging in the filter F160W band using HST WFC3. The main lensing galaxy is denoted as G, whereas the main perturbers within 5\arcs of the lens are numbered from P1 to P9.}
    \label{fig:wfi_hst}
\end{figure*}

\section{Time-delay measurement}
\label{sec:TD}

\subsection{Optical photometric monitoring}
 As one of the first discovered strongly lensed quasars, multiple attempts at measuring the time delay of \HEonze with different observation campaigns have been published in the literature. First, \citet{wisotzki98} published 18 spectroscopic observations with the 3.6m ESO telescope spanning over 5 years and estimated that the time delay between images A and B, \dtab, should be between 109 and 292 days. Reanalysis of these data by \citet{gilmerino02} and \citet{pelt02} showed that this sampling is insufficient for a precise delay. With a better sampling of the R-band photometry between 1997 and 2006 with the OGLE and SMARTS programs\footnote{\url{http://www.astro.yale.edu/smarts/}},  \citet{poindexter07} estimated $\dtab = 152 \pm 3$ days using a Legendre-polynomial fitting technique. Later, \citet{morgan08} re-estimated to $\dtab = 162.2 \pm 6.1$ days by jointly fitting the time delay and accretion disk size of the quasar by exploiting the microlensing signal contained in the light curves. However, the method employed depends on physical assumptions about the quasar accretion disk profile and population of microlenses that could bias the measurement. 

In this work, we used three new unpublished datasets and applied the established \pycs spline-fitting method to re-estimate the time delay of \HEonze in a data-driven way without relying on any prior assumption about physical properties of the system.
The first dataset is provided by the SMARTS telescope, which monitored it from 2003 to 2016 with a $\sim$ weekly cadence. Next, the COSMOGRAIL program \citep{courbin05} observed it every $\sim$ 4 days with ECAM and C2 at the Leonhard Euler 1.2m Swiss Telescope from 2013 to 2018 and during the 2017 season with WFI at the MPG/ESO 2.2m Telescope every $\sim$ 2 days. These three datasets are called SMARTS, ECAM, and WFI, respectively.

 \begin{figure*}
 

\sidecaption
\includegraphics[width= 0.7\textwidth]{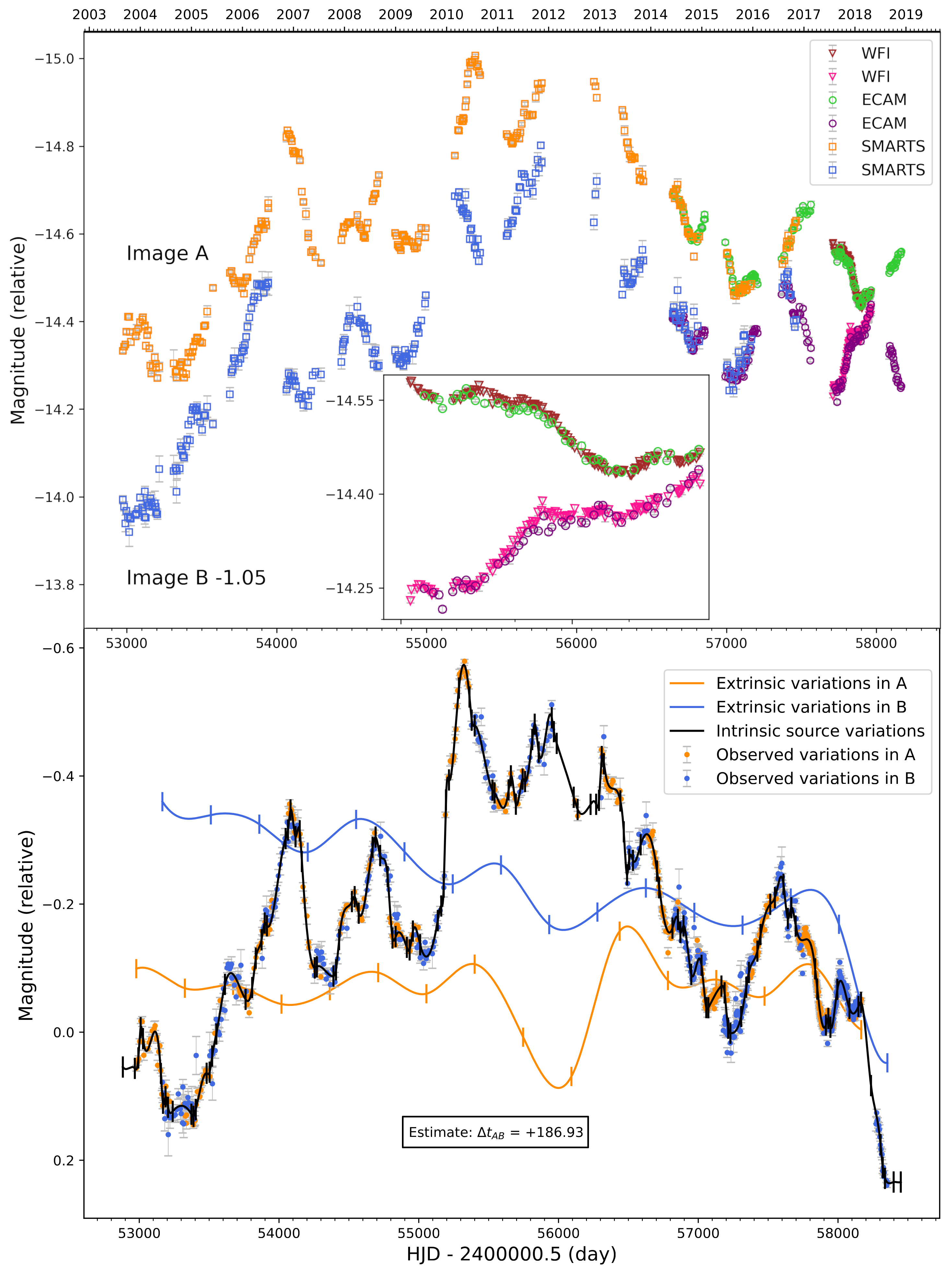}
\caption{\toppanel \HEonze R-band light curve obtained by joining three datasets: SMARTS (yellow and blue squares), ECAM (purple and green circles), and WFI (brown and pink triangles). For clarity, the light curve of B was shifted by -1.05 mag. The inset zooms on the WFI dataset span, showcasing its superior sampling to the ECAM dataset. \bottompanel Example of a simultaneous fit of an intrinsic light curve with $\eta = 45$ days and $n_{\rm ml} = 15$. The time shift obtained by this fit gives a point estimate of the system's time delay. By repeating such measurements with randomised starting points 800 times, we obtained the time-delay measurement of this ($\eta$, $n_{\rm ml}$) configuration. \label{fig:lcs_data}}
 
\end{figure*}

For each dataset, the quasar images' photometry was measured with the same methodology as \citet{millon20b} that we briefly summarise here. 
After bias, flat, and sky corrections, each exposure's PSF is modelled as a Moffat profile fit combined with regularised pixel adjustments of five stars with similar luminosity as the lens in the field. The MCS deconvolution algorithm \citep{magain98} is then applied to the lensed images in order to deblend the flux of an image from its counterpart, the lensing galaxy, and the arc. 

Finally, a deconvolution of reference stars is carried out using the previously built PSF. Stars displaying the most photometric stability are chosen to calculate a median photometric normalisation coefficient for each exposure. This step helps to correct image-to-image systematics resulting from PSF variations across epochs. Fig.~\ref{fig:wfi_hst} shows a stack of the exposures taken with the WFI, with the stars used to compute the PSF and the normalisation of each epoch.

The resulting light curves are shown on the top panel of Fig.~\ref{fig:lcs_data}. The three monitoring datasets presented are merged into a single light curve by fitting for possible magnitude and flux shifts between instruments on the overlapping parts of the light curves. The resulting combined dataset is referred to as 'ECAM+SMARTS+WFI'. 

\begin{figure*}
\sidecaption
 \minipage{0.35\textwidth}
     \includegraphics[width =\linewidth] {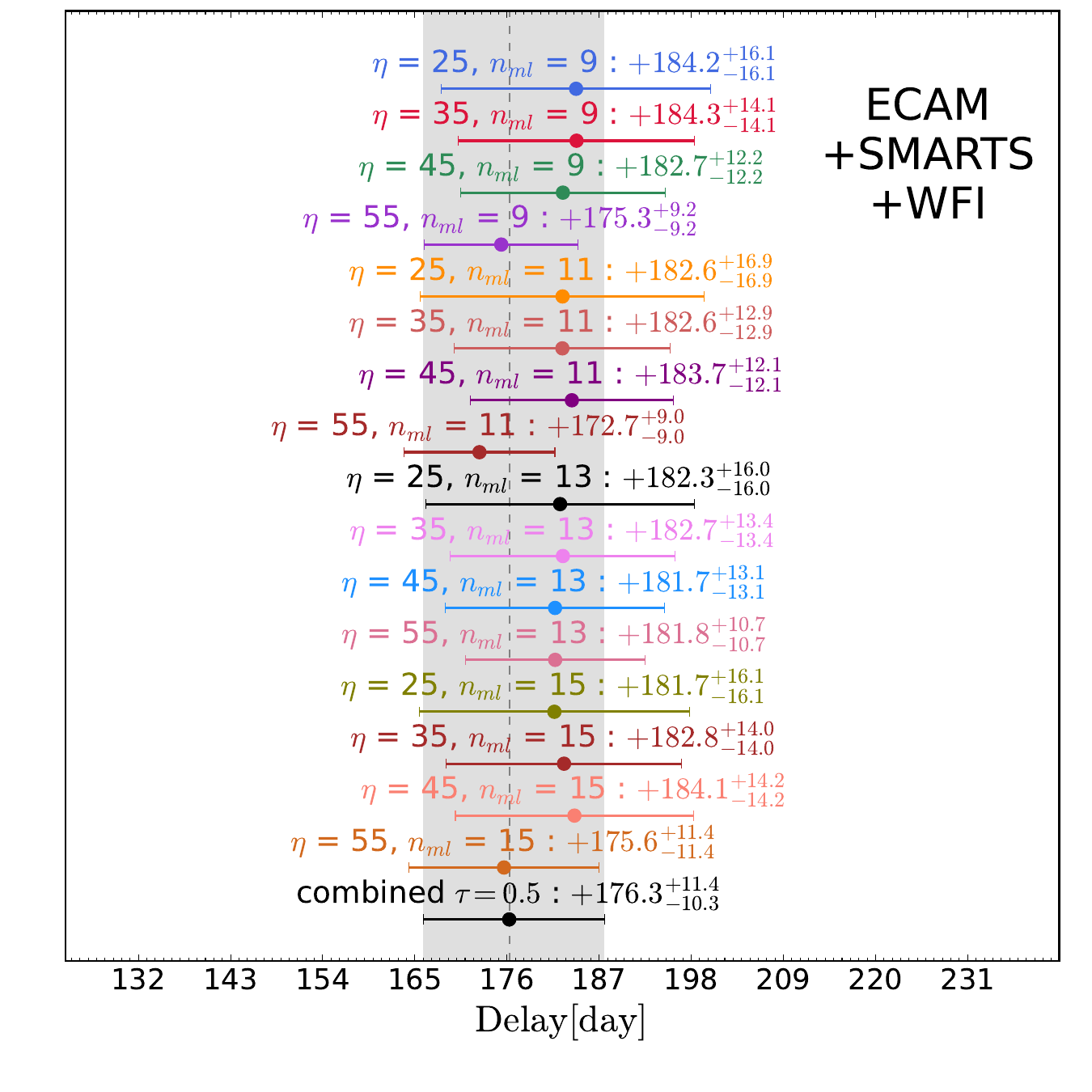}
 \endminipage
\minipage{0.35\textwidth}
\vspace{0.3cm}
     \includegraphics[width =\linewidth] {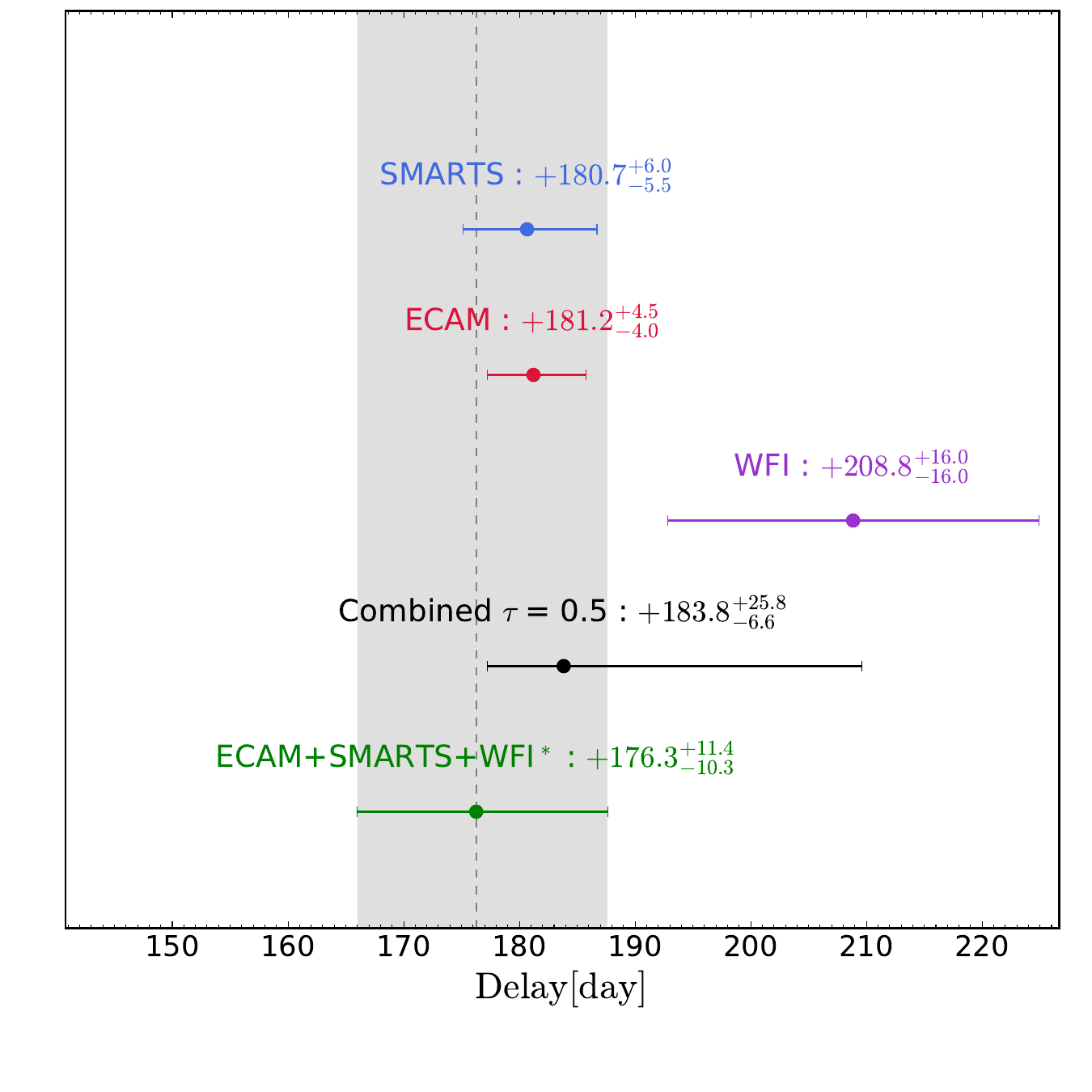}
 \endminipage
  \caption{\leftpanel Measurements of the time delay for different configurations of ($\eta$,  $n_{\rm ml}$) with the merged ECAM+SMART+WFI dataset.  \rightpanel Measurement of \dtab \ for the different datasets. Since the datasets overlap, they cannot be considered fully independent. We therefore use our estimate on the joint dataset ECAM+SMARTS+WFI (green data point) as our final estimate. In both panels, the 'combined $\tau = 0.5$' value was obtained by marginalising over the measurements listed above it. }
    \label{fig:marginalisationspl}
\end{figure*}

\subsection{Curve shifting procedure}
The time delay was computed using the method described extensively by \citet{millon20}, relying on free knot spline fitting implemented in the PyCS python package\footnote{We used version 3.0 from \url{https://cosmograil.gitlab.io/PyCS3/}} \citep{millonjoss}. Free-knot splines are piece-wise polynomials where the mean number of days between two consecutive knots is controlled by the parameter $\eta$, which filters out variation with shorter time scales. A single free-knot spline is used to model the intrinsic variability of the quasar, which is identical in both light curves; meanwhile, an additional spline is fitted on each light curve to model the extrinsic variability caused mainly by microlensing. The number of degrees of freedom is controlled by the hyperparameter $\eta$ for the intrinsic spline. For the extrinsic splines $n_{\rm ml}$ sets the number of knots throughout the whole curve. We fitted the intrinsic, image A and B extrinsic components, and the time delay between the light curves. A visual first guess of the time delay is used as a starting point of the fit \footnote{This visual measurement is done on the publicly available D3CS tool: \url{https://obswww.unige.ch/~millon/d3cs/COSMOGRAIL\_public/}}. The bottom panel of Fig.~\ref{fig:lcs_data} shows an example of such a fit, giving us a point estimate of the time delay. 

We repeated the fit 800 times with starting points randomised uniformly in a range of 10 days around the first guess value, we took the median value as the time delay measurement for a given ($\eta$, $n_{\rm ml}$). 

To assess the uncertainty of this measurement in a data-driven way, a set of mock light curves is generated with built-in time delays uniformly drawn in an interval of $\pm 10 $ days around the median value obtained previously. we used the generative noise model described by \citet{millon20} to ensure that the simulated light curves have the same noise properties as the real data. The distribution of the difference between the measurement and the truth for each of the mocks gives an estimation of the systematic uncertainty (mean of the distribution) and the statistical uncertainty (standard deviation) of the time delay estimation. The final uncertainty of the time delay estimation is the sum in quadrature of these two sources of uncertainty.

The choice of values for $\eta$ and $n_{\rm ml}$ is guided by the fact that quasars vary significantly faster than microlensing. Moreover, whether a given feature is mathematically attributed to the intrinsic or extrinsic splines depends on the flexibility balance between the splines. We therefore reproduced the same measurement for different values of these two hyperparameters. Following \citet{millon20}, we took $\eta \in [25,35,45,55] $ days and $n_{\rm ml}\in [9,11,13,15]$ days when fitting the SMARTS, ECAM, and the merged datasets. In contrast, the microlensing in the much shorter WFI dataset is fitted with a spline having either one or two knots ($n_{\rm ml}\in [1,2]$) since the microlensing variability is expected to alter the light curve on time scales longer than a year.

The left panel of Fig.~\ref{fig:marginalisationspl} shows the time-delay measurements obtained for each hyper-parameter configuration ($\eta$, $n_{\rm ml}$) considered. 
The measurements obtained are consistent with each other with variable precision. We note that for a given $n_{\rm ml}$, fits with $\eta=55$ are slightly lower, which shows that the intrinsic fit is marginally too rigid. We are therefore conservative by keeping this value and should not use higher values for $\eta$. We gradually marginalised the most precise measurement with the others in order of increasing precision until the tension with the rest of the sample is lower than the threshold $\tau$. Following the methodology of \citet{millon20}, we used $\tau=0.5$ (i.e. no individual measurement is further than 0.5$\sigma$ away from the final measurement) to have a conservative uncertainty estimation without being biased by outlier measurements.

Since the three variability components are degenerate and not physically informed, the resulting splines cannot be interpreted as the true source and microlensing variations. However, a visual inspection shows that the magnitude scales of the extrinsic variations are below $0.3$ mag over several years, consistent with the expected microlensing behaviour \citep[e.g.][]{mortonson,schneider06, mosquera11}. We, however, note that the number of features in the extrinsic splines is higher than in most systems, which hints towards a more complex microlensing behaviour, as observed by \citet{schechter03} in previously published light curves, making the extrinsic fit sensitive to the hyperparameter changes. 

The same experiment is carried out separately for each dataset, and the right panel of Fig.~\ref{fig:marginalisationspl} shows the time delay measured in each case, the combined value, and the measurement on the merged ECAM+SMARTS+WFI dataset. The marginalisation over the three datasets would be relevant if each dataset were subject to different sources of systematic errors in the photometric measurement. However, the photometry was computed with the same method for each dataset, and no discrepancies were observed between time delays measured with ECAM and WFI data on other lens systems \citep[e.g.][]{millon20, millon20b}. As shown by the lower panel of Fig.\ref{fig:lcs_data}, the extrinsic variations are complex during the 2017 season. The particular case of WFI is challenging because it covers only 253 days: given the long time delay, the light curves overlap on less than 100 days.  Even though distinct features are visible and well-sampled in the WFI data, the complexity of the microlensing noticed previously cannot be adequately modelled with a single season. 
The time delay with the WFI dataset is, thus, not robust enough for the marginalised time delay value to be trusted. Therefore, we preferred to use the ECAM+SMARTS+WFI dataset, which allowed us to use the entire period of the light curve to constrain the microlensing behaviour as well as to take advantage of the high cadence and high signal-to-noise (S/N) display of the WFI dataset.

We did not take into account a possible microlensing time delay \citep{tie17} because it impacts the time delay only by a day in the worst cases \citep{bonvin18}, which is negligible given the length of the time delay. Therefore, we use the time-delay $\dtab=176.3^{+11.4}_{-10.3}$ days, i.e. a 5.8$\%$ precision for our cosmography analysis.

\section{Spatially resolved kinematics of the lens }

\subsection{Integral field unit spectroscopy with MUSE \label{sec:MUSE}}

A 660-second IFU observation of the system was carried out with the multi-unit spectroscopic explorer (MUSE) instrument \citep{Bacon2010} on March 13th 2019, in wide-field mode with adaptive optics (PI: A. Agnello, program 0102.A-0600(A) and 106.215F.001). The observation wavelength ranges from 4700.03 \AA \ to 9351.28 \AA \ with a spectral resolution of R $\sim$ 2500 corresponding to an instrumental dispersion of $\sigma_{\rm inst} = 120\ \kms$. The spatial resolution is 0.2\arcs pixel$^{-1}$ and the mean seeing throughout the wavelength after the adaptive optic correction is $\sim 0.57\arcs$. The data reduction has been carried out as in \citet{sluse19}, using the MUSE reduction pipeline.

\begin{figure}[ht!]
\centering
        \includegraphics[width=\linewidth]{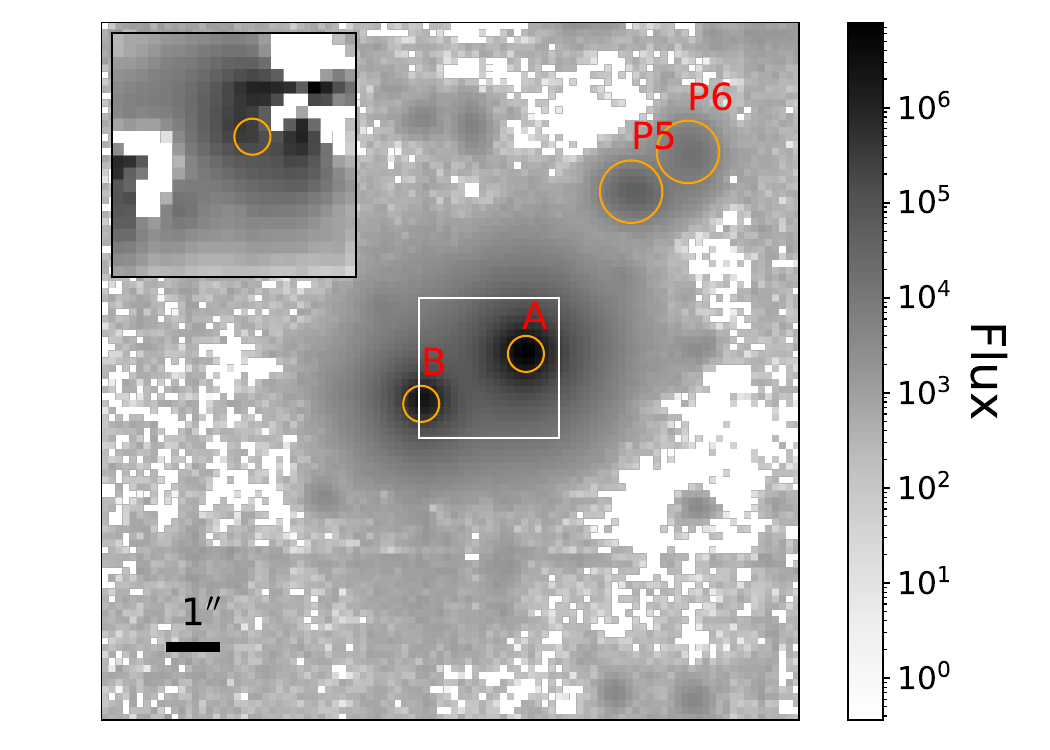} 
    \caption{MUSE data cube summed over all wavelengths. The inset displays the point-source subtracted cube, revealing the lens galaxy light. The apertures used to extract individual spectra of the quasar images, lens centre, and perturbers shown in Fig.~\ref{fig:spectra} are represented by orange circles.  \label{fig:musecube}}
    
\end{figure}

This range includes the lens redshifted lines \ion{Ca}{H \& K} lines at 3933 \AA \ and 3968 \AA \ along with the G-band at 4304 \AA \ , while 5806 \AA \ to 5965 \AA \ wavelengths were cut out due to the notch filter.  
The data cube integrated across wavelength is shown in the top panel of Fig.~\ref{fig:musecube}. To ease spectral deblending of the quasar images from the lens galaxy, we subtracted the point source light at each wavelength of the data cube. To do so, at each wavelength, the point sources and lens galaxy light are fitted simultaneously with a Moffat and S\'ersic profile with relative positions fixed to the ones measured on the HST data.  
For each wavelength frame, the Moffat part of the fit is then subtracted from the data to obtain the data cube displayed in the inset of Fig.~\ref{fig:musecube}.  

The spectrum of the centre of the lens galaxy obtained from this subtracted cube is shown in Fig.~\ref{fig:spectra} and displays observation with an S/N of 30$ \AA^{-1}$, which is used to measure \slos.
We also retrieved new high-quality quasar spectra thanks to a small aperture integration on images A and B, and the spectra of some perturbers annotated on Fig.~\ref{fig:wfi_hst} (See Appendix \ref{appendix:muse_spec}). Finally, we extracted the spectra of the two brightest perturbers (P5 and P6) with a sufficiently high S/N ($\sim 11 \ \AA^{-1}$ and $\sim 7 \ \AA^{-1}$ respectively) to measure their redshifts. 

\begin{figure*}[h!]
\centering
  \includegraphics[width=\linewidth,trim=2 2 2 2,clip]{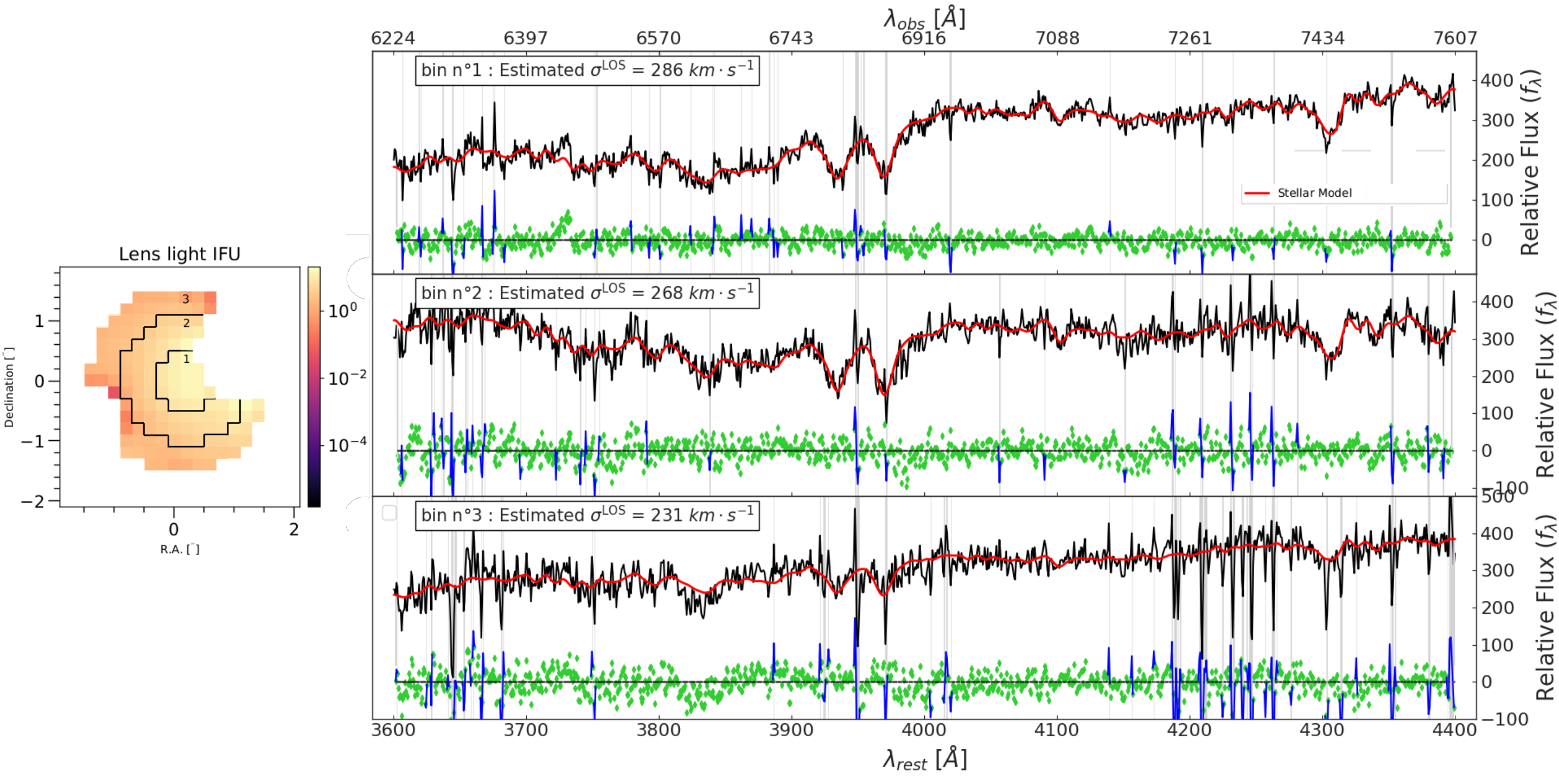}

\caption{Velocity dispersion point-estimate in the three radial bins of \HEonze's lens galaxy. \leftpanel Mean of the MUSE data cube after PSF subtraction and masking with overlaid bin numbers and contours. \rightpanel Example of a pPXF fit of the integrated spectra of each bin, where data masked for the fit are marked in grey. \label{fig:vdispestimate}}
\end{figure*}

\subsection{Stellar template fitting \label{sec:vdispmeas}}

As shown by the left panel of Fig.~\ref{fig:vdispestimate}, we masked the pixels from the lens light cube further than 1.8\arcs~from the centre of the galaxy, as well as an area of 0.8\arcs~around the position of images A and B. In order to maximise the number of constraints on the radial velocity profile, the frame is then divided into three concentric rings of 0.53\arcs~width with S/N from 14 $\AA^{-1}$ to 30 $\AA^{-1}$. 
The integrated spectra of each bin are then fitted using the penalised pixel fitting (pPXF) method\footnote{We used version 8.2 from \url{https://pypi.org/project/ppxf/}} \citep{cappellari17, Cappellari2023}. This method models the galaxy spectrum with a weighted linear combination of stellar spectra, broadened by a convolution with the galaxy line-of-sight velocity distribution. Additive Legendre polynomials are included in the model to improve the robustness to template mismatch and dust reddening. Furthermore, a multiplicative polynomial accounts for spectral flux calibration mismatches between data and templates. Any contaminating signal from residuals of the quasar images blended with the galaxy is accounted for by including a scaled quasar spectrum presented in Fig.~\ref{fig:spectra}.

As shown in Fig.~\ref{fig:spectra}, the most constraining features of the lens galaxy spectrum for the \slos \ measurement are the \ion{Ca}{H}, \ion{Ca}{K}, and G-band absorption lines identified respectively at 6800 \AA, 6828 \AA \, and 7395 \AA \ in the observed frame. To focus the fit on these lines and avoid regions of the spectrum that could be dominated by the \ion{C}{iv} and \ion{C}{iii} emission of the quasars, we restricted the wavelength range to [6220, 7780] \AA \ (in the rest-frame [3600, 4400] \AA). 

Following \citet{knabel25}, we used three stellar template libraries: Indo-US \citep{valdes04}, MILES \citep{sanchez06,falcon11}, and X-shooter spectral library \citep{verro22}, cleaned from incomplete or low-quality spectra and convolved them down to the same FWHM as the observation's Line Spread Function in the galaxy's rest frame, estimated at $1.46\AA$ with the empirical relation given by \citet{bacon23}.

An exploratory set of fits allowed us to determine that an additive polynomial of degree 6 combined with a multiplicative polynomial of degree 0 yields the most robust measurements across the different template libraries.

An example of the \slos \ estimate in each bin given by a fit with this fiducial setup is shown in Fig.~\ref{fig:vdispestimate}. 

We then combined the fits from different template libraries using the bayesian information criterion \citep[BIC, ][]{schwarz78}.
It is defined as:
\begin{equation}
    BIC = k \cdot ln(n_{\rm d}) -2 ln(\hat{L}),
\end{equation}
where $\hat{L}$ is the maximised value of the likelihood function of the model $\displaystyle M$, i.e. $\displaystyle {\hat {L}=p(d\vert {\widehat {\xi_{\rm max}}})}$, where $\displaystyle{{\widehat {\xi_{\rm max}}}}$ are the parameter values that maximise the likelihood function, $n_{\rm d}$ is the number of data points (i.e. the number of unmasked spectral pixels), $k$ is the number of parameters in the model. 

The central value and statistical uncertainties are given by the BIC-weighted mean of the measurements obtained with the different libraries, while the systematic uncertainty is computed as the weighted-mean difference between the central value and each library measurement.

To test the robustness of the measurement against potential contamination of the \ion{Ca}{K} and G-band by sky emission lines, we repeated the measurement while masking the first  ([3954;4000] \AA) or the latter ([4275;4325] \AA). We obtained measurements in excellent agreement with the original one and therefore use this one for the cosmographic analysis.

The final measurement, along with the corresponding covariance matrix, is shown in Table~\ref{tab:slos}.

\begin{table}[]
    \centering
    \caption{\slos measurements and covariance matrix of each bin. }
    \label{tab:slos}
    \begin{tabular}{c|c|c c c}
    \hline
     & \textbf{Mean value [\kms]} & \multicolumn{3}{c}{\textbf{Covariance matrix}}\\ [0.6 ex]
         & & $\slos_1$& $\slos_2$& $\slos_3$ \\ [0.6 ex]
         \hline
        
       $\slos_1$  & 278$\pm$ 7 (sys) $\pm$ 13 (stat) & 241 &  &  \\[0.7 ex]
       $\slos_2$  & 259 $\pm$ 9 (sys) $\pm$ 8 (stat)& 46 & 419 &  \\ [0.7 ex]
       $\slos_3$  & 222 $\pm$ 8 (sys) $\pm$ 29 (stat) & -6 & -59 & 946 \\ [0.7 ex]
       \hline
    \end{tabular}

\end{table}

\section{Line-of-sight analysis \label{sec:los}}
\subsection{Wide-field imaging}

\begin{figure}
    \centering
    \includegraphics[width=\linewidth]{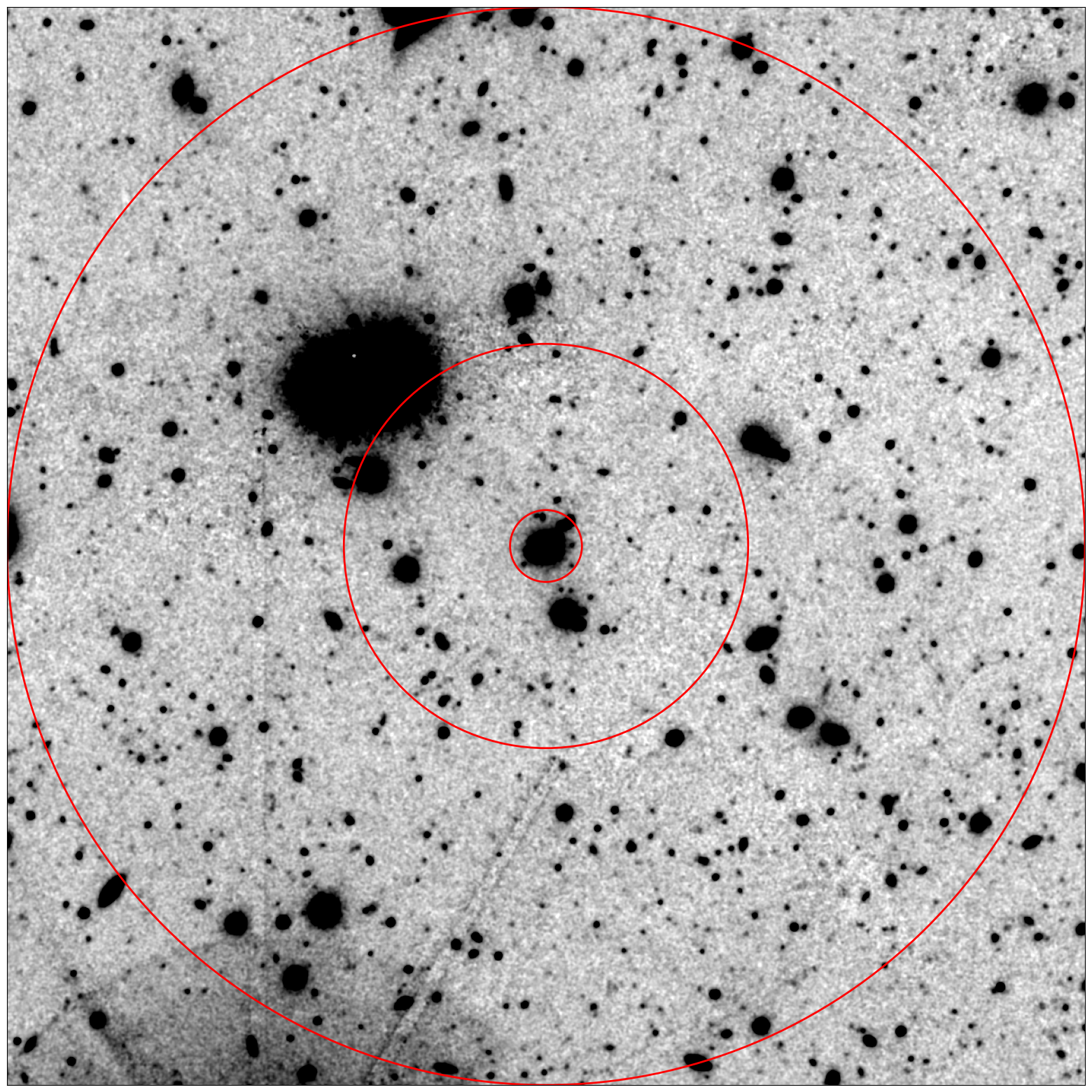}
    \caption{Wide-field imaging of \HEonze in the $i$-band. Red circles denote regions within 8\arcs, 45\arcs\, and 120\arcs of the centre of the lens.}
    \label{fig:widefield}
\end{figure}

The \HEonze field was targeted by a multi-band imaging campaign to assess the contribution of line-of-sight structures to the lensing signal.  The data acquired included $u$-band with a total exposure time of $t_{\rm exp} = $20706~s with Megacam on the Canada France Hawaii Telescope (Program ID: 14AT01; PI Suyu); Subaru SuprimeCam imaging in the $g$ ($t_{\rm exp} = $600~s), $r$ ($t_{\rm exp} = $4800~s), and $i$ ($t_{\rm exp} = $600~s) bands (Program ID: o14220; PI Fassnacht, the latter is shown in Fig. \ref{fig:widefield}) ; and Subaru MOIRCS near-infrared imaging in the $J$ ($t_{\rm exp} = $3600~s), $H$ ($t_{\rm exp} = $5760~s), and $Ks$ ($t_{\rm exp} = $2940~s) bands (Program ID: o15202; PI Fassnacht).  For the current analysis of the line-of-sight, we only considered the SuprimeCam imaging.

The SuprimeCam data were reduced in a two-phase process.  The initial steps utilised the SDFRED reduction package to do the overscan and bias correction, flat-field correction, distortion and atmospheric distortion corrections, and masking of some bad regions.  The second phase used custom Python-based scripts to mask bad pixels and to combine the individual exposures.  

The SuprimeCam imager provides a field of view of approximately 30$^\prime$ on a side, which is substantially larger than what is needed for the analysis presented in Sect.\ref{sec:kextmeas}.  We therefore used only a 4$^\prime \times 4^\prime$ section centred on the lens system to generate the galaxy catalogues.  The catalogues were produced by running {\tt SExtractor} \citep{bertin96} in dual-image mode, using the deep $r$-band data as the detection image and the $i$-band image as the measurement image.  The photometric zeropoints were determined by comparing the instrumental magnitudes of stars in the field to the magnitudes measured by SDSS for those stars.

\begin{figure*}
    \sidecaption
    \includegraphics[width = 0.7\linewidth]{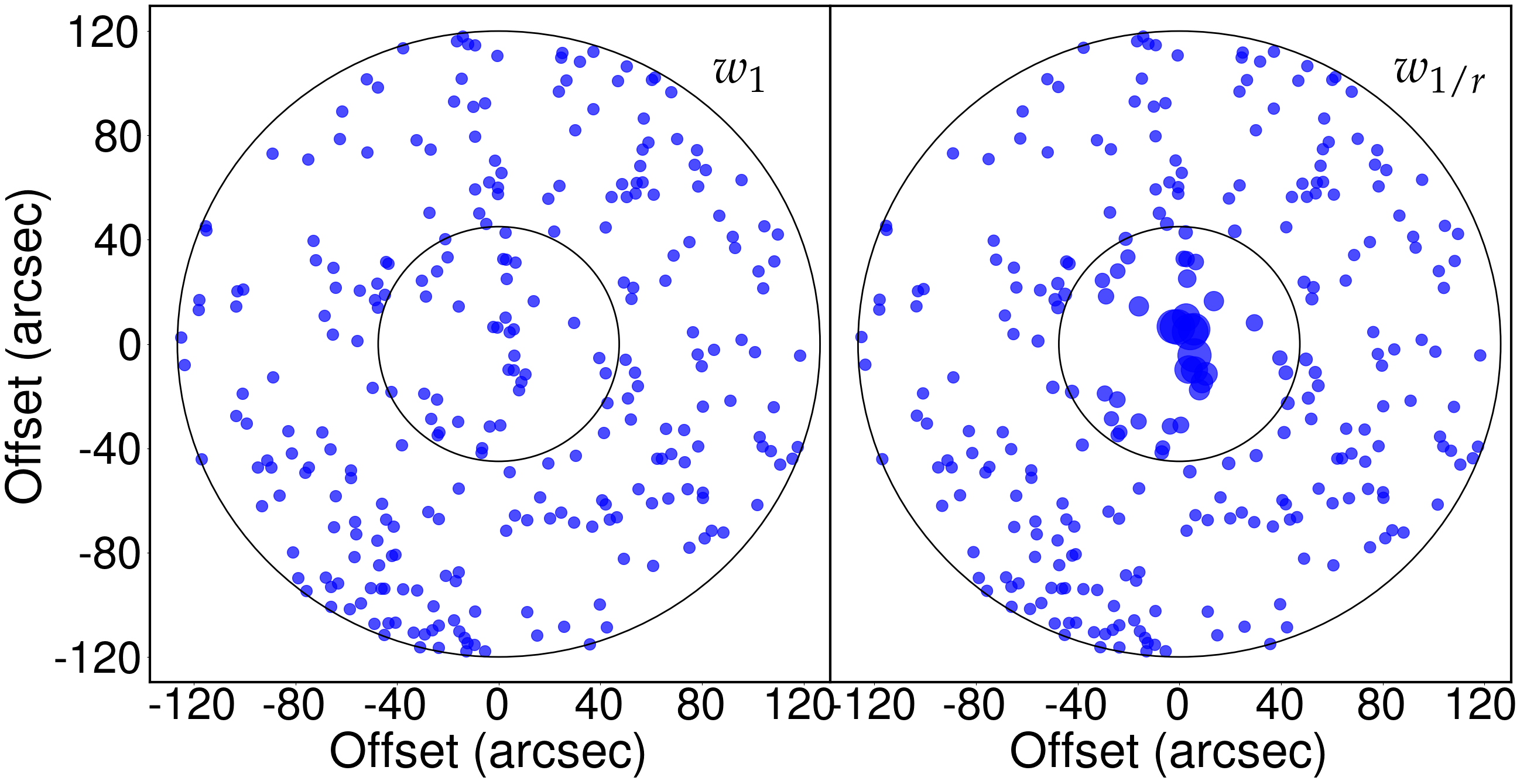}
    \caption{
Position and weight (size of the dot) of galaxies between us and the source. Black circles indicate the 45\arcs and 120\arcs apertures. Blue dots correspond to objects with $i$-magnitudes less than 24.}
    \label{fig:catweights}
\end{figure*}

\begin{figure}
 \centering
        \centering
        \includegraphics[width=\linewidth]{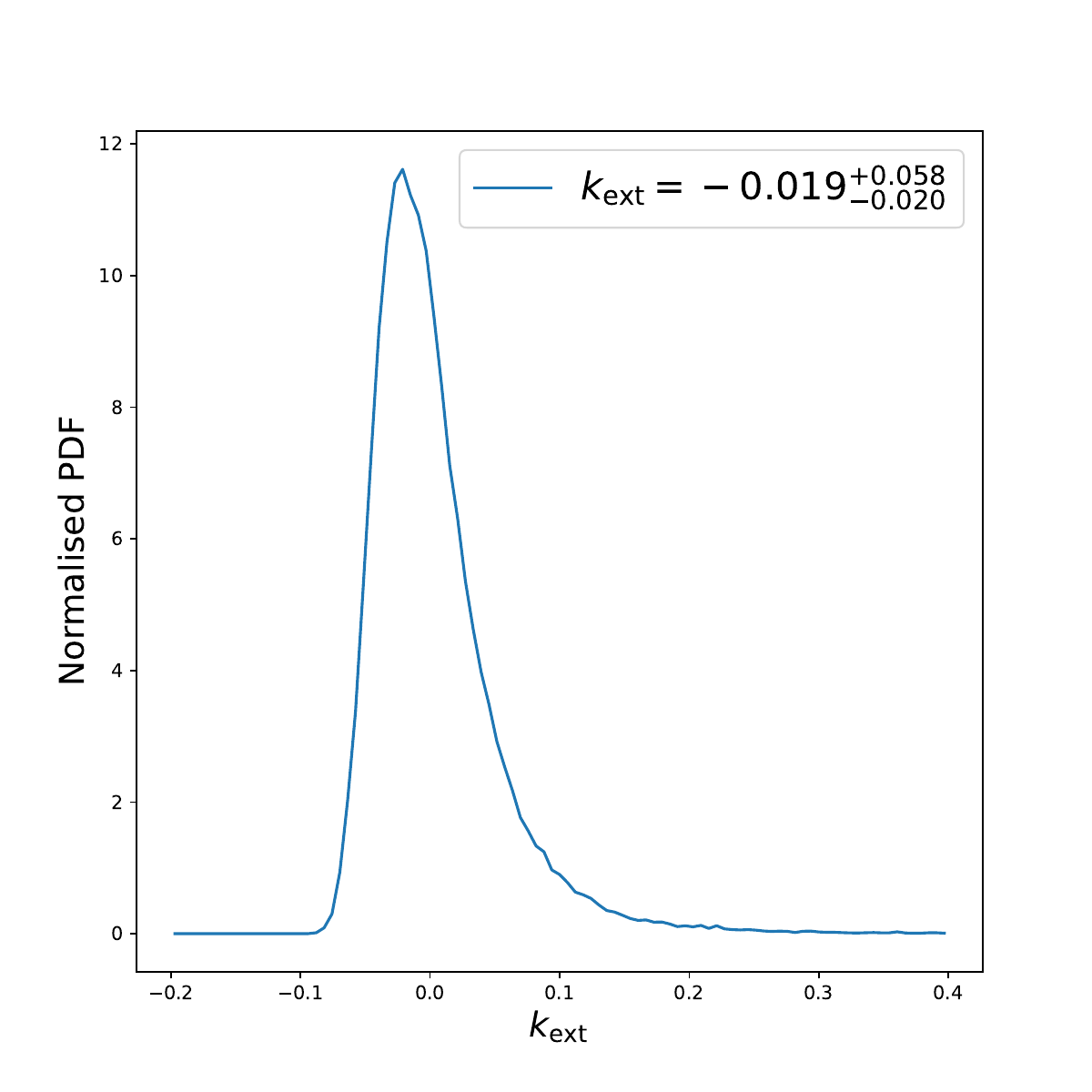}
    \caption{
    \kext \ measurement showing that the line-of-sight of \HEonze is along a slightly underdense region of the Universe. Therefore, the fact that multiple perturbers are near the main lens is coincidental and the approximation $\kext \approx \kappa_{\rm s}$ holds.}
    \label{fig:kextmeas}
\end{figure}

\subsection{Perturbers and groups explicitly included in the lens model \label{sec:pert}}

As shown in \fig\ref{fig:wfi_hst}, we identified nine perturbers within 5\arcs of the lens in the HST imaging. Our initial step, hence, involved identifying the perturbers that we need to explicitly include to the lens model.

The `flexion shift' was introduced by \citet{mccully17} to determine which line-of-sight galaxies should be explicitly included in the lens modelling and which ones should be implicitly accounted for through the computation of $\kappa_{\rm ext}$.

The specific influence of line-of-sight objects depends on whether the dominant-lens approximation holds, indicating that the critical density of these objects is much smaller than that of the primary deflector \citep{fleury21}.  

The difference in lens image position induced by a given nearby galaxy is given by:
\begin{equation}
    \Delta_3 x = f\left(\beta\right) \frac{\left(\theta_{\rm E,lens}\cdot \theta_{\rm E,pert}\right)^2}{\theta^3},
    \label{eq:flexionshift}
\end{equation}
with $\theta_{\rm E,lens}$ and $\theta_{\rm E,pert}$ as the Einstein radii of the lens and perturber, and $\theta$ as the angular separation of the lens and perturber. The function $f\left(\beta\right)$ equals 1 if the perturber is between the lens and the observer, while $f\left(\beta\right) = \left(1-\beta\right)^2$ if the perturber is behind the main lens. $\beta$
 is the pre-factor of the lens deflection in the multi-plane lens equation: $\beta = \left(D_{\rm dp} D_{\rm s}\right) / \left(D_{\rm p} D_{\rm ds}\right)$. \citet{mccully17} found that perturbers with $\Delta_3 x \leq 10^{-4}$ are accurately accounted for with a global external convergence term, whereas perturbers with a higher flexion shift should be individually modelled with a multi-plane treatment.
 
We used the $M_* - \slos$ relation obtained \citet{zahid16} on galaxies up to redshift 0.7 to estimate the mass-to-light ratio, M/L, of the P5 perturber using its velocity dispersion \slos$_{\rm P5}$ measured with the MUSE data (see Appendix \ref{app:slosp5} for more details). The flux determined with a circular aperture around P5 in the F160W band converted into luminosity using a fiducial cosmology with $\hc=70$ \kmsmpc and $\Omega_{\rm m} =0.3$. By applying the same M/L ratio to the other perturbers, we estimated their mass and \slos. Given the uncertainty of the \slos$_{\rm P5}$ measurement, the uncertainty on the fiducial cosmological parameters and the fact that some perturbers are assumed to be further than the range studied by \citet{zahid16}, can be neglected. 
By approximating the perturbers' mass distributions with an SIS, we computed their Einstein radius, $\theta_{\rm E,SIS}$, with :

\begin{equation}
    \theta_{\rm E,SIS} = 4\pi\left(\frac{\slos}{c}\right)^2 \frac{D_{\rm ps}}{D_{\rm s}}
\end{equation}
with $D_{\rm ps}$ and $D_{\rm s}$ the angular diameter between the source and perturber, and between the observer and the source, respectively. 

We assumed that the perturbers without a redshift measurement lie between the closest perturber, P5 and the source. We therefore repeated the computation with redshifts uniformly drawn between $z_5$ and $z_{\rm s}$, and took the mean, the 16th, and 84th percentile ($1-\sigma$ in a Gaussian distribution) of the resulting distribution as the final estimate.

We could then estimate the flexion shift caused by each perturber using Eq.~\ref{eq:flexionshift}. The results presented in Table~\ref{tab:perturbergroups} show that even when marginalising on their redshift, the flexion-shifts of these perturbers are all potentially larger than $10^{-4}$. They can therefore significantly influence the image's position and should be included in the mass model.

In addition, the subtraction of the PSF, lens, and source light with a preliminary SIE model revealed the presence of a luminous component $\sim 0.6 \arcs$ from image B. Without a counter-image indicating that this feature belongs in the source, we treated this component as a perturber, hereafter P9. Because of its faintness, the spectra of P9 could not be extracted from the MUSE data cube, and its mass could not be estimated with the same technique employed for all other perturbers.

\renewcommand{\arraystretch}{1.3}
\begin{table*}[h!]
\begin{flushleft}
    \caption{Galaxies and groups of galaxies identified as potential perturbers.}
    \label{tab:perturbergroups}
    \begin{tabular}{c|c|c|c|c|c|c}
    \thead{Perturber \\ or Group} & \thead{Number of \\ galaxies} & \thead{R.A} & \thead{Dec} & \thead{Redshift} & \thead{$\theta_{\rm E,SIS}$ [\arcs]} &  \thead{log($\Delta_3 x$)}  \\[0.6ex]
    \hline
    
     P1 & 1 & 166.64076 & -18.35770& (0.729) & $0.2^{+0.5}_{-0.2}$ & $-4.2^{+0.5}_{-1.9}$\\[0.6ex]
     P2 & 1 & 166.64030  &  -18.35620 & (0.729) & $0.2^{+0.6}_{-0.2}$ & $-3.1^{+0.5}_{-1.9}$\\[0.6ex]
     P3 & 1 & 166.63993  & -18.35468 & (0.729) & $0.1^{+0.1}_{-0.1}$ & $-4.5^{+0.6}_{-1.4}$\\[0.6ex]
     P4 & 1 & 166.63954 &  -18.35476 & (0.729) & $0.3^{+0.3}_{-0.3}$ & $-4.1^{+0.4}_{-1.2}$ \\[0.6ex]
     P5 & 1 & 166.63818 &  -18.35526 & 0.3575 & $0.37^{+0.16}_{-0.16}$ & $-3.0^{+0.5}_{-0.5}$\\[0.6ex]
     P6 & 1 & 166.63770 & -18.35497 & 0.505 & $0.35^{+0.26}_{-0.27}$ & $-3.3^{+0.4}_{-1.3}$\\[0.6ex]
     P7 & 1 & 166.63786 & -18.35677 & (0.729) & $0.2^{+0.7}_{-0.2}$ & $-4.2^{+0.5}_{-1.8}$\\[0.6ex]
     P8 & 1 & 166.63765 & -18.35777 & (0.729) & $0.3^{+0.4}_{-0.3}$ & $-3.7^{+0.5}_{-2.0}$ \\[0.6ex]

     \hline
     \end{tabular}
\end{flushleft}
     \begin{flushleft}
     \begin{tabular}{c|c|c|c|c|c|c|c|c|c}
     \hline
   & & & & & & & R$_{\rm 200}$ [\arcs] & $v_{\rm disp}$ [\kms] & R$_{\rm s}$[\arcs] \\[0.6ex]
         Group 1 &  \ \ \ \ \ \ \ 6 \ \ \ \ \ \ \ & 166.63123 & -18.34910 & \ 0.441 \ \ &
          \ \ 3.9$^{+3.7}_{-1.9}$ \ \
          & -3.4$^{+1.3}_{-1.1}$ &
          $194^{+52}_{-70}$
          & $383^{+94}_{-117}$ &$26.2^{+8.0}_{-9.4}$\\[0.6ex]
         Group 2 & 11 & 166.63579 & -18.36816 & 0.495 &
         2.4$^{+1.7}_{-0.9}$
         
         & -3.2$^{+1.3}_{-0.9}$ & 
         $148^{+33}_{-132}$
         & $314^{+83}_{-280}$&$20.9^{+6.9}_{-19.6}$\\[0.6ex]
         Group 3 & 7 & 166.62309 & -18.36606 & 0.506 &
         2.8$^{+2.6}_{-1.2}$
         
         & -3.5$^{+0.6}_{-0.6}$ & 
          $163^{+49}_{-49}$
          & $354^{+93}_{-97}$ &$24.3^{+8.0}_{-7.8}$\\[0.6ex]
         Group 4 & 7 & 166.62614 & -18.34243 & 0.577 &
         3.7$^{+3.4}_{-1.7}$
         & -3.4$^{+0.9}_{-0.8}$ & 
         $183^{+30}_{-45}$
         & $418^{+85}_{-109}$ &$30.4^{+7.6}_{-9.3}$\\[0.6ex]
         Group 5 & 7 & 166.63055 & -18.34246 & 0.578 &
         2.7$^{+2.4}_{-1.1}$
         & -3.5$^{+0.8}_{-0.7}$ &$153^{+30}_{-45}$
         & $359^{+68}_{-88}$ &$25.3^{+5.9}_{-7.3}$\\[0.6ex]
         Group 6 & 7 &166.63574 & -18.34827 & 0.580 &
         2.6$^{+1.4}_{-0.9}$
         
         & -2.8$^{+0.9}_{-0.9}$ & 
          $152^{+30}_{-45}$
          & $334^{+88}_{-98}$&$23.1^{+7.6}_{-8.0}$\\[0.6ex]
    \end{tabular}
    \end{flushleft}
    \tablefoot{Top table: Properties of the perturbers identified on the HST imaging within 5\arcs of the lens. Einstein radii and flexion shifts based on their luminosity and the M/L ratio computed with P5. Redshifts in parentheses are assumed. Bottom table: Characteristics of the groups of galaxies with log($\Delta_3 x$) > -4.}
    
\end{table*}

\subsection{Identification of galaxy groups in the vicinity of the lens}

Our search for galaxy groups along the line-of-sight relies on a catalogue of spectroscopic redshifts of field-of-view galaxies. To compile that catalogue, we have obtained multi-object spectroscopy of galaxies up to $\sim$2 arcmin using the FORS2 instrument mounted on the Antu VLT telescope (PID: 092.0515, PI: D. Sluse), and using GMOS at the Gemini-South Telescope (PID: GS-2013B-Q-28; PI: T.Treu). The observational setup, data reduction, and target selection strategy are similar to the one described in \citep{sluse17}. Individual spectra typically cover the range 4500-9200 $\AA$ for FORS2 data and 4400-8200 $\AA$ for GMOS. Within a 30$^{\prime\prime}$ radius around the lens, we also extracted spectra of the sources present in the MUSE data used for kinematics (Sect.~\ref{sec:vdispmeas}. We used 1D cross-correlation methods to measure redshifts: the \texttt{xcsao} task implemented in IRAF was used for the FORS data, and MARZ \citep{Hinton16} was used for the other data sets. This yields an ensemble of 131 reliable unique redshifts (96 with FORS2, 18 with GMOS, 17 with MUSE) distributed typically in the range z $\in [0.04, 0.8]$. We complemented the catalogue with redshift measurements from \cite{Momcheva15} who targeted galaxies as faint as $I\sim 21$\, mag within a radius of 15 arcmin of \HEonze. This sample adds 387 unique redshifts to the catalogue, out of which 47 are within a radius of 6 arcmin of the lens, i.e. the region where we expect groups to have first-order contribution to the main lens gravitational potential.

The identification of galaxy groups is done using the algorithm described in \citet{sluse17}, \citet{sluse19}, and \citet{buckleygeer20}, and we refer the reader to these publications for details. This algorithm is based on the group-finding algorithms of \citet{wilman05} and \citet{ammons14}. For this measurement, we have computed the group centroid from the luminosity-weighted positions of the group members.  The groups likely to perturb the modelling of the main lens are summarised in Table \ref{tab:perturbergroups}. Groups 4, 5, and 6 highlight the group finder's sensitivity to initial conditions and interlopers, with some galaxies appearing in multiple groups. We therefore included them separately in the lens models (see Sect. \ref{sec:models}).
In Fig.~\ref{fig:groups}, we show, for each identified group, the right ascension and declination of the accepted and rejected member galaxies as well as the distances and velocities relative to the final group centroid.

\subsection{External convergence (\kext) measurement \label{sec:kextmeas}}
Similar to the previous time-delay cosmography studies \citep[e.g.][]{rusu17, birrer18, rusu20}, we used the number count technique to determine \kext. This quantity determines the contribution of objects not included in the lens modelling. 
Number counts involve measuring the galaxy number density near the lens as a summary statistic and comparing it to reference fields. This comparison helps determine whether the LOS is over- or under-dense compared to the average background \citep[e.g.][]{fassnacht11,greene13,wells23}.

It can be summarised in four main steps:
\begin{itemize}
    \item Lines of sight (typically several arcminutes wide squares) are drawn from a large comparison field and compared to the LOS of the lens. Modern survey datasets covering hundreds to thousands of square degrees ensure a sufficiently large comparison field to avoid sampling bias. The photometric catalogues of the fields are cleared from objects more distant than the lensed source and fainter than the used magnitude cut on the $i$ < 24 \citep[e.g. ][]{fassnacht11}. This cut is designed to consider only objects with reliable photometry. \\
    \item The weight value of a given LOS $W_{ i}$ is then computed as the ratio of the weighted number counts for galaxies in the lens field to the same statistic in the reference field.
\begin{equation}
    W_{ i} = \frac{\Sigma_{ j} w_{ j,{\rm lens}}}{\Sigma_{ j} w_{ j,i}}
\end{equation}
    where $j$ indexes the galaxies in a given LOS and $w$ is the weight of a galaxy. Different types of weights can be considered \citep[e.g. ][]{greene13,rusu17,wells23} such as the galaxy's distance to the center ($1/r_{ j}$), its potential ($m_{ j}/r_{ j}$), its redshift ($z_{ j} (z_{\rm s} - z_{ j})$) or simply be the same for each galaxy ($w_{ j} = 1$). The weight value obtained, therefore, acts as a tracer of the external convergence of the lens LOS. \\
\item Similar weight values are computed in simulated fields with known \kext \citep{hilbert09}. Previous works \citep[e.g.][]{rusu17, wells23}, have shown that using the Millennium Simulation yields a reliable estimate of \kext \ because it provides catalogues of galaxies for several values of the external convergence at sufficient points to represent the Universe accurately. Since the weight value relies on ratios, many dependencies on the simulation's underlying cosmological parameters are expected to cancel out. \\
\item The posterior probability of \kext \ can hence be computed as follows 
\begin{equation}
    p(\kext\vert d) = \int p_{\rm sim}(\kext\vert W_i)p(W_i\vert d)\Pi_{ i} dW_{ i} .
\end{equation}

Here, $p(W\vert d)$ represents the probability distribution of the weighting scheme $W$ given the data, and $p_{\rm sim}(\kext\vert $W$)$ denotes the probability distribution of \kext in the simulated dataset, conditioned on a specific value of the 
weight $W$. \citet{greene13} showed that the combination of multiple evaluations relying on different weights (while adequately taking into account covariances between 
estimates) significantly improves the precision and robustness of the measurement.
\end{itemize}

As detailed previously, the treatment of perturbers within 5\arcs\ from the lens is part of the explicit lens model. We, therefore, masked this region from the number counting and removed the galaxies forming the perturber groups presented in Table \ref{tab:perturbergroups}, as in \citet{wells23}, we limited the field to 120\arcs around the lens. The weights used are the number counts $w_{1}$ and the inverse distance $w_{1/r}$ of objects. Their computation is displayed in Fig.\ref{fig:catweights} and \fig \ref{fig:kextmeas} displays the final measurement $\kext \ = -0.019_{-0.020}^{+0.058}$. 
Our procedure, in fact, determines $\kappa_{\rm s}$, the convergence up to the source which we used as a proxy for \kext. The external convergence is technically computed using the observer-deflector and deflector-source convergences ($\kappa_{\rm d}$ and $\kappa_{\rm ds}$) through $\kext = \left(1-\kappa_{\rm d}\right)/\left(1-\kappa_{\rm s}\right)\cdot\left(1-\kappa_{\rm ds}\right)$. However, \citet{tang25} showed that ignoring this correction affects the \hc measurement only when the LOS up to the deflector is significantly over- or under-dense, which is not the case here. At the population level, the expected bias on \hc due to this approximation is about 0.1\%, which is negligible compared to the present 4.6\% precision of TDCOSMO25.

Therefore, this measurement is used for the rest of the cosmographic analysis, without further correction, in compliance with the perturber inclusion strategy.

\section{Lens modelling}
\label{sec:models}

\subsection{HST imaging}

In the context of H0LiCOW \citep{suyu17}, Program 12889 (P.I: S.H Suyu) took 12 exposures of 36 seconds along with 24 exposures of 599 seconds of \HEonze with wide-field camera 3 (WFC3) in the F160W\footnote{Data in different filters is also available, but since the extended source is not visible on those, they don't bring any additional constraint to the lens modelling} band. A 4-point dither pattern with parallelogram shapes having half-pixel shifts was applied in order to reach a resolution of 0.08\arcs per pixel. 
The imaging displayed in the left panel of Fig.~\ref{fig:wfi_hst} showcases the main challenges for modelling this system. First, the point source images are particularly bright with the PSF wings of both images aligned with the centre of the galaxy. Accurate modelling of the PSF's wings is, therefore, crucial to model the light of the lens. Secondly, the quasar host's light is very dim and distinguishable only after proper subtraction of the lens light (see Fig.~\ref{fig:decomppl}). This makes the lens light fitting a primordial and sensitive step to properly reconstruct the shape of the arc, a key constraint to mass modelling. Thirdly, the system's environment is particularly crowded
with eight luminous galaxies within 15$^{\prime\prime}$ of the main lens.
The system is also characterised by a large and asymmetric separation of images A and B positions from the lens centre of 1.17\arcs and 2.05\arcs, respectively.  

\subsection{Setup and workflow}

 \begin{figure*}
    \centering
    \includegraphics[width=\linewidth]{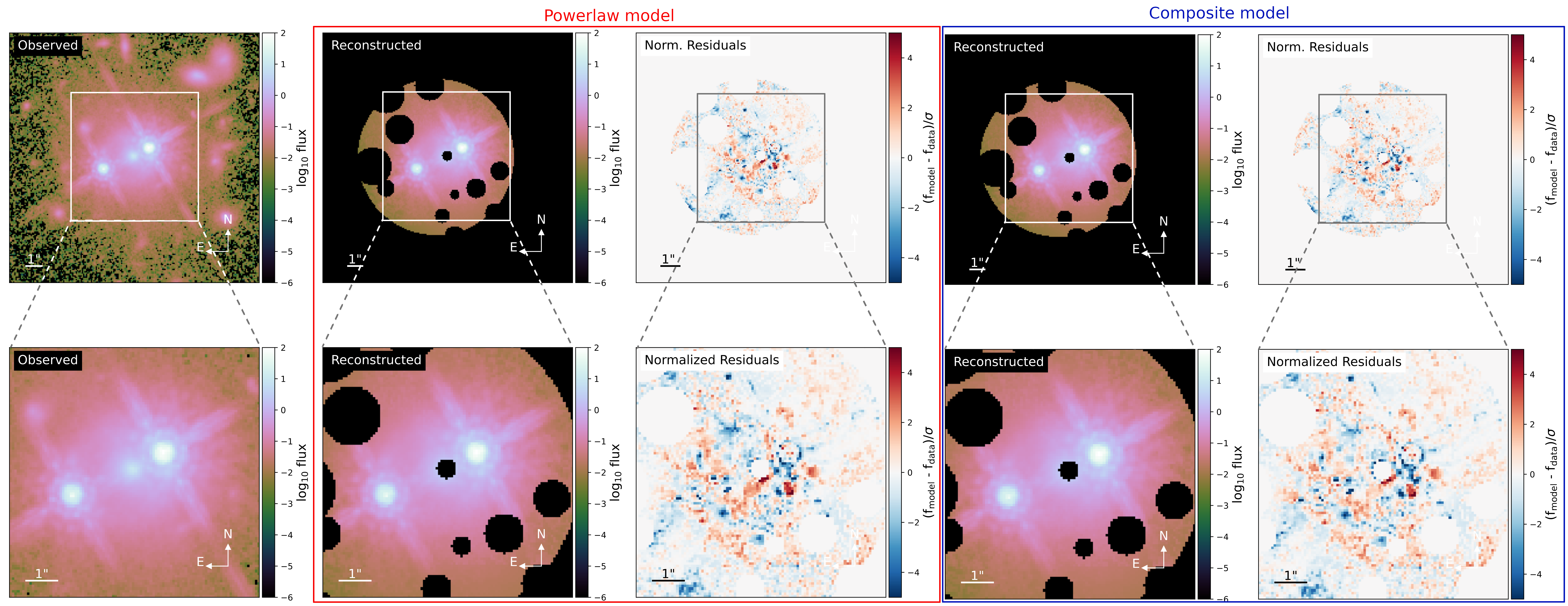}
    \caption{Reconstructed image of \HEonze using the best power-law and composite models. From left to right: The imaging data, reconstructed image, and normalised residuals are displayed for the power law model (PEMD + Shear + perturbers 2, 5, 6, 9 + groups 1, 2, 3, and 4) and composite models (NFW + Double chameleon + Shear + perturbers 2, 5, 6, 9 + groups 1, 2, 3, and 4). }
    \label{fig:residuals}
\end{figure*}

\begin{figure*}
    \centering
    \includegraphics[width=\linewidth]{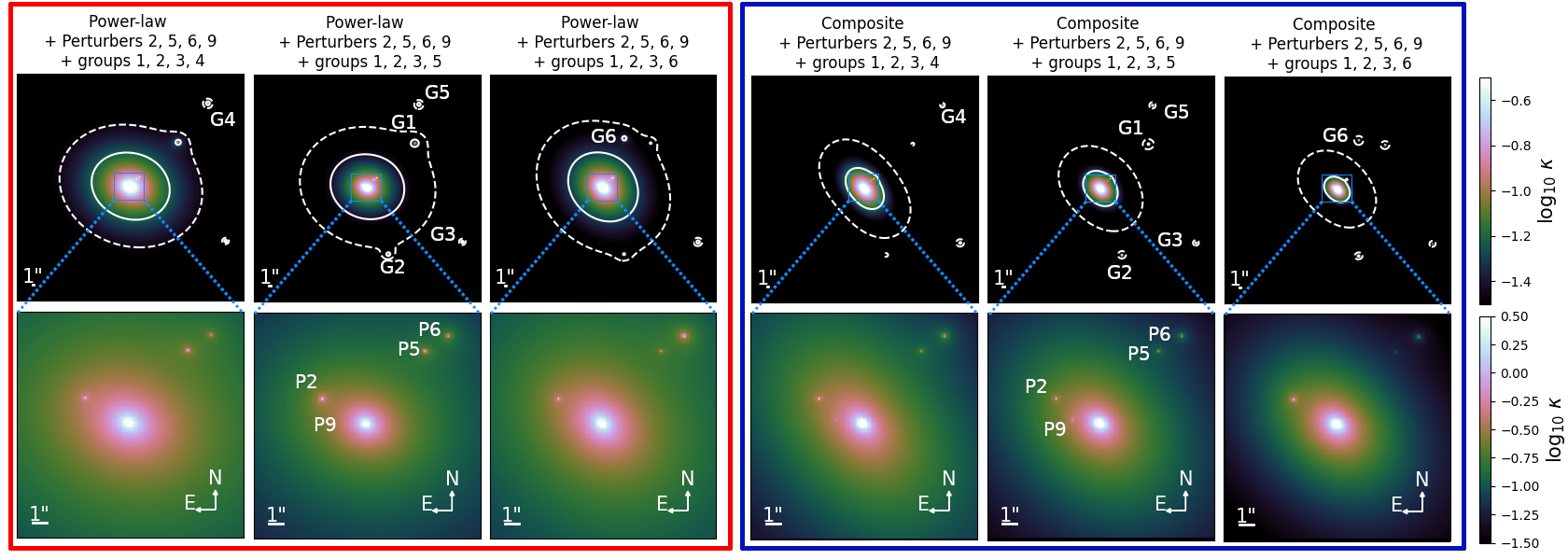}
    \caption{\HEonze modelled convergence $\kappa$ with the two families of models and different perturbers and groups inclusion strategies. The top row shows a wide-field including all groups and the bottom row zooms in on the region closer to the main lens; white lines indicate the area enclosing 25, 50, and 75\% of the total mass.}
    \label{fig:convflexmag}
\end{figure*}

\subsubsection{Main deflector mass and light model:}
Following the methodology of the most recent TDCOSMO lens models \citep[e.g. ][]{shajib20,shajib22}, we modelled the lens galaxy with two families of models: power law and a composite alternatively.

In the first case, the mass is parametrised as a power-law elliptical mass distribution (PEMD) for which the convergence at the position $\left(\theta_1,\theta_2\right)$ in the frame aligned with the major and minor axis of the deflector is given by 
\begin{align}
    \kappa_{\rm PL}\left(\theta_1,\theta_2\right) = \frac{3-\gamma_{\rm pl}}{2}\left(\frac{\theta_{\rm E}}{\sqrt{q_{ m}\theta_1^2+\theta_2^2/q_{ m}}}\right)^{\gamma_{\rm pl}-1} ,
    \label{eq:pemd}
\end{align}
where $\theta_{\rm E}$ is the Einstein radius, $\gamma_{\rm PL}$ the slope of the profile, and $q_{ m}$ its axis ratio. This convergence map is then rotated by the position angle $\phi_{ m}$ to belong to the on-sky coordinate frame.

The lens light is modelled by a superposition of two S\'ersic profile which gives the light intensity at the position $\left(\theta_1,\theta_2\right)$ following 
\begin{align}
    I_{\rm{Sersic}}\left(\theta_1,\theta_2\right)=I_{\rm eff}\exp \left\{-b_{ n}\left[\left(\frac{\sqrt{\theta_1^2+\theta_2^2/q^2}}{\theta_{\rm eff}}\right)^{ 1/n}-1\right]\right\}
    \label{eq:sersic}
\end{align}
with $\theta_{\rm eff}$ the effective radius (i.e. the product average of semi-major and semi-minor axis), also defined as the half-light radius, thanks to the normalising factor $b_{ n}$, $I_{ \rm eff}$ the intensity at $\theta_{ \rm eff}$. $q$ is its axis ratio, and $n$ the S\'ersic index. 

The positions of the two Sersic profiles are tied together throughout the fit, and the position of the mass component's centre is fixed on the light component.

The chameleon profile can be parametrised as the difference between two non-singular
isothermal ellipsoid (NIE) profiles \citep[e.g. ][]{suyu14} 
\begin{align}
    \kappa_{\rm Chm}(\theta_1, \theta_2) = \frac{\kappa_{0,{\rm Chm}}}{(1+q)} \left[ \frac{1}{\sqrt{\theta_1^2+\theta_2^2/q^2 + 4w_{\rm c}^2/(1+q)^2}}- \right. \nonumber\\ 
    \left. \frac{1}{\sqrt{\theta_1^2+\theta_2^2/q^2 + 4w_{ t}^2/(1+q)^2}} \right] 
    \label{eq:chameleon}
\end{align}

where $q$ is the axis ratio and $w_{\rm c}$ and $w_{\rm t}$ are sizes of the cores of the two NIE with $w_{\rm t}$ > $w_{\rm c}$.

For the composite model, the baryonic mass 
is described by a double chameleon profile (i.e. two superposed chameleon profiles presented in  Eq.\ref{eq:chameleon}). 

The dark matter halo is modelled with an elliptical Navarro-Frenk-White (NFW) profile, whose volumic density is given by 
\begin{equation}
    \rho(r) = \frac{\rho_0}{\left(\frac{r}{R_{\rm s}}\right) \left(1 + \frac{r}{R_{\rm s}}\right)^2}
    \label{eq:nfw}
\end{equation}
where $\rho_0$ is a normalisation and $R_{\rm s}$ is the scale radius.

Following the measurement of \citet{gavazzi07} on galaxies that have a similar \slos \ as the one measured in this system, we expect the NFW's scale radius to be within $R_{\rm s} =58\pm 8$ kpc. To prevent any prior bias, we took this value as a starting point within a wide uniform prior $R_{\rm s} \sim \mathcal{U}(25,100)$kpc.

We used a double chameleon profile to model the light, and the fitted parameters are fixed to the parameters of the baryonic mass model. 

To avoid the lens light modelling from being biased by background light while still predicting accurate flux at the position of the arc, we masked pixels further than 4.3\arcs from the lens. We adjusted the mask locally to mask the light of all the perturbers and masked the lens's central pixels because the light is not fitted perfectly in this region. Masking the few pixels close to the centre of the lens prevents the model from predicting source flux at this position, which can possibly bias the inferred mass profile. 

\subsubsection{Source light model}
The visible source light is smooth, and we modelled it with a single elliptical S\'ersic profile in both families of models. Contrary to previous TDCOSMO studies, no further structure is detected in the lensed arc. Thus, it is not required to add shapelets to describe extra complexity.

\subsubsection{Perturbers' model}
To include the perturbers in the mass model, we added an SIS component to the mass model for each perturber. Even though we did not fit their light in the final model, a preliminary S\'ersic fit gives us their position. Since P5 and P6 have different redshifts than the lens, we used the multi-lens-plane formalism introduced in Eqs. \ref{eq:multiplane} and \ref{eq:fermatpotmultiplane}. For the other perturbers considered, we assumed that they lie in the same plane as the lens. We considered the perturbers P2, P5, and P6 as they have the highest flexion shifts, and P9 because of its proximity to image B. 
Furthermore, the perturber groups listed in Table \ref{tab:perturbergroups} are parametrised with NFW profiles. We estimated their scale radii using their virial mass (based on measured $R_{\rm 200}$ and $v_{\rm disp}$) and estimation of the NFW concentration parameter, $R_{\rm 200}/R_{\rm S}$ from \citet{duffy08}. The groups 4, 5, and 6 are alternatively included in the model.

\subsubsection{PSF modelling \label{sec:psf}}
To complete the light model, each component is convolved by the PSF. As a first guess of the PSF, five stars of the field were stacked together using the PSFr software \citep{birrer22}.

Since the spectrum of the stars in the field used for the PSF initial guess does not match the quasar one, we finalised the PSF model with four successive repetitions of the following sequence with the {\tt Lenstronomy} software \footnote{We used version 1.9 from \url{https://github.com/lenstronomy}} \citep{birrer18}:
\begin{enumerate}
    \item mass and light model parameter optimisation using {\tt Lenstronomy}'s built-in particle swarm optimiser routine \citep[e.g. ][]{eberhart95,birrer18}
    \item iterative reconstruction of the PSF on lens and source subtracted residuals. We also adjusted the PSF noise map in a radius of 0.5\arcs around its centre to adapt the constraining power of these pixels. 
\end{enumerate}

\subsubsection{Posterior sampling and combination of different models \label{sec:lensmodel_posteriorsampling}}

After obtaining a satisfactory PSF, shown in Fig.~\ref{fig:psf}, we used {\tt Lenstronomy}'s built-in MCMC routine relying on EMCEE \citep{foreman13} to sample the light and mass parameters ($\xi_{\rm light}$ and $\xi_{\rm mass}$) posterior by maximizing the likelihood expressed as  

\begin{equation}
    \mathcal{P}( \xi_{\rm light},\xi_{\rm mass} \vert {\rm d}) \propto \mathcal{L}(d \vert \xi_{\rm light},\xi_{\rm mass} ) p(\xi_{\rm light},\xi_{\rm mass}),
\end{equation}
where $\mathcal{L}(d \vert \xi_{\rm light},\xi_{\rm mass} )$ is the likelihood of the parameters given the data and $p(\xi_{\rm light},\xi_{\rm mass})$ is the prior on the parameters. During this step, the data used to constrain the posterior are the unmasked pixels of the HST observation $\mathcal{D}_{\rm Img}$ and the time delay, \dtab. The likelihood maximised by the sampling is therefore computed with 
\begin{align}
        \mathcal{L}(d\vert \xi_{\rm light},\xi_{\rm mass} ) = \mathcal{L}( {\rm Img} \vert \xi_{\rm light},\xi_{\rm mass} ) \nonumber \\ 
        + \mathcal{L}(\dtab \vert \xi_{\rm light},\xi_{\rm mass} )
\end{align}
which is, up to an additive constant, 
\begin{align}
     = \log\left[-\frac{1}{2}\sum_{ i}^{N_{\rm pix}} \frac{(\mathcal{D}_{\rm Img, i}- M_{\rm Img, \textit{i}})^2}{\sigma_{ i}^2}\right]
     + \log\left[-\frac{1}{2}\frac{(\dtab - M_{\Delta t_{\rm AB}})^2}{\sigma_{\Delta t_{\rm AB}}}\right],
\end{align}

where $M_{\rm Img}$ and $M_{\Delta t_{\rm AB}}$ are the model-predicted image and time delay of the system; $\sigma_{ i}$ is the pixel's noise, and $\sigma_{\dtab}$ the uncertainty on the time delay.
We used ten times as many walkers as there are parameters in the model, and we used the last 1\,000 out of 20\,000 iterations to construct the posterior of a given modelling setup. The best reconstructions of the HST imaging with power-law and composite models shown in Fig.~\ref{fig:residuals} demonstrate that all relevant features in the light are correctly predicted.
The convergence profile obtained by each model family in every perturber-inclusion scenario is displayed in Fig.\ref{fig:convflexmag}. We notice that, with both model families, the convergence of the perturbers P5, P6, and G1 varies substantially between the 
different scenarios. This can be attributed to the fact that these components are aligned in the same direction relative to the main lens, resulting in a degenerate effect on the system. Conversely, the convergence of the perturbers P2, P9, and groups G2, G3 are more stable as they are more azimuthally spread.
We also note that in every scenario, the composite model is cuspier than the power-law one.

The BIC values computed with the maximum likelihood of each model are presented in Table~\ref{tab:bic}. The variance of the BIC is $\sigma_{ \Delta BIC}$ = 73 within the power-law models and $\sigma_{ \Delta BIC}$ = 48 within the composite models.  

Similarly to \citet{birrer19} and \citet{shajib22}, we determined the weight of each posterior distribution following 
\begin{equation}
    \mathcal{W} = \frac{1}{\sqrt{2\pi}\sigma_{ \Delta BIC}}\int ^\infty_{-\infty} f(x) \rm{exp}\left(-\frac{(BIC-x)^2}{2\sigma_{ \Delta BIC}} \right) dx,
\end{equation}
where $f(x)$ is the evidence ratio function given by 

    \begin{align}
  f\left(x\right) \equiv 
  \begin{cases}
    1 & \textrm{if $x \leq$  BIC$_{\rm min}$,}  \\
    \rm{exp}(\rm{BIC}_{\rm min}-$x$) & \textrm{if $x$ > BIC$_{\rm min}$,} \\
  \end{cases} 
\end{align}
with BIC$_{\rm min}$ the BIC of the reference model with the minimum BIC value. 

To create a single posterior out of $N$ multiple models, we summed all the posteriors using their normalised weights 
\begin{equation}
    \mathcal{W}=  \frac{\sum_{n}\mathcal{W}_{ n} p(\xi_{\rm light},\xi_{\rm mass}\vert d)}{\sum_{n} \max(\mathcal{W}_{ n})}.
    \label{eq:weight}
\end{equation}

\begin{figure}
    \centering
    \includegraphics[width = \linewidth]{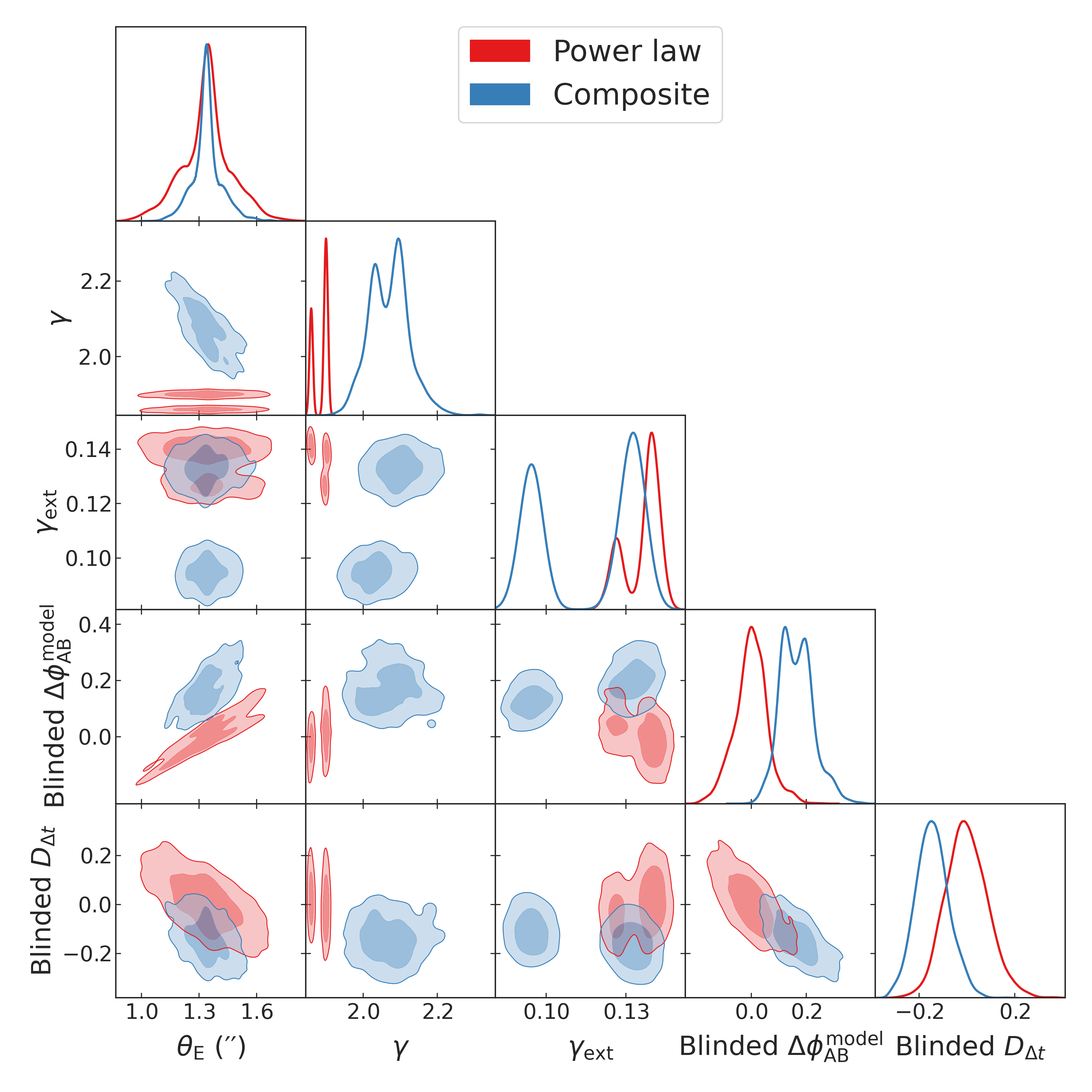}
    \caption{BIC weighted comparison of the power-law and composite models. For each family, we marginalised over the three posteriors given by the three perturber inclusion scenarios, which results in the multi-modality of some of the parameters such as $\gamma$. We obtained $\theta_{\rm E}^{\rm PL} = 1.356 \pm 0.115$, $\gamma_{\rm pl} = 1.88\pm 0.04$, $\gamma_{\rm ext}^{\rm PL} = 0.14\pm 0.01$, $\theta_{\rm E}^{\rm Comp} = 1.37 \pm 0.057$, $\gamma_{\rm ext}^{\rm Comp} = 0.12\pm 0.02$, and $\gamma^{\rm Comp} =2.03 \pm 0.08$   
    \label{fig:cornerplvscomp}}
\end{figure}

This procedure is done separately for the composite and power law models, and we compare the results in Fig.~\ref{fig:cornerplvscomp}. For both models, the Einstein radius, the slope, and the time-delay distance are robust to the different perturber configurations, whereas the external shear shows multi-modality because the G6 group centre is azimuthally significantly different from groups G4 and G5 (see Fig. \ref{fig:convflexmag}). A blind comparison of the two modelling families allowed us to see that a 2.1$-\sigma$ disparity between the two families on $\gamma$ induces a 1.3$-\sigma$ difference in the Fermat potential and 1$-\sigma$ for the time-delay distance between the two families.
The external shear we inferred for \HEonze\ is higher than in other TDCOSMO lenses. As shown by \citet{etherington23}, model-predicted \gext in power-law + shear configurations is often overestimated even in perturber-free systems because it is degenerate with potential azimuthal structures, which have a negligible impact on \hc at the population level\citep[][]{vandevyvere22}. 
In Section \ref{sec:pert}, we considered only nine perturbers, out of which five are explicitly modelled, but many more are visible in the right panel of Fig.\ref{fig:wfi_hst}. These perturbers are expected to induce external shear on the main lens and induce a large \gext. More marginally, the positions of the centre of mass of the groups of pertubers, G1-G6, are poorly constrained and might also inflate the value of \gext.

\begin{table}
    \centering
    \caption{Summary of lens modelling scenarios and their relative weight within each family. 
    \label{tab:bic}}
\begin{tabular}{c|c|c|c}
    Lens model & Perturbers included & BIC & Relative Weight \\
    \hline
    \hline
     Power-law & \thead{P5, P6, P2, P9, \\  groups 1, 2, 3, 4 }& 9875 & 0.39 \\
     Power-law & \thead{P5, P6, P2, P9, \\  groups 1, 2, 3, 5} & 9878 & 0.22 \\
     Power-law & \thead{P5, P6, P2, P9, \\  groups 1, 2, 3, 6 } & 9945 & 0.39 \\
     
     \hline
     Composite & \thead{P5, P6, P2, P9, \\  groups 1, 2, 3, 4 } & 9781 & 0.36\\
     Composite & \thead{P5, P6, P2, P9, \\  groups 1, 2, 3, 5 } & 9865 & 0.36\\
     Composite & \thead{P5, P6, P2, P9, \\  groups 1, 2, 3, 6 } & 9783 & 0.28 \\

\end{tabular}
\tablefoot{These posterior serve as priors for the kinematic fitting.}
\end{table}

\section{Cosmographic inference \label{sec:inference}}

We combined the mass and light posteriors obtained with the lensing and time delay constraints with the resolved kinematics measured in Section \ref{sec:vdispmeas} to measure \dd, \ddt, and \hc. 

\begin{figure*}
\sidecaption
\minipage{.3\textwidth}
    \includegraphics[width=\linewidth]{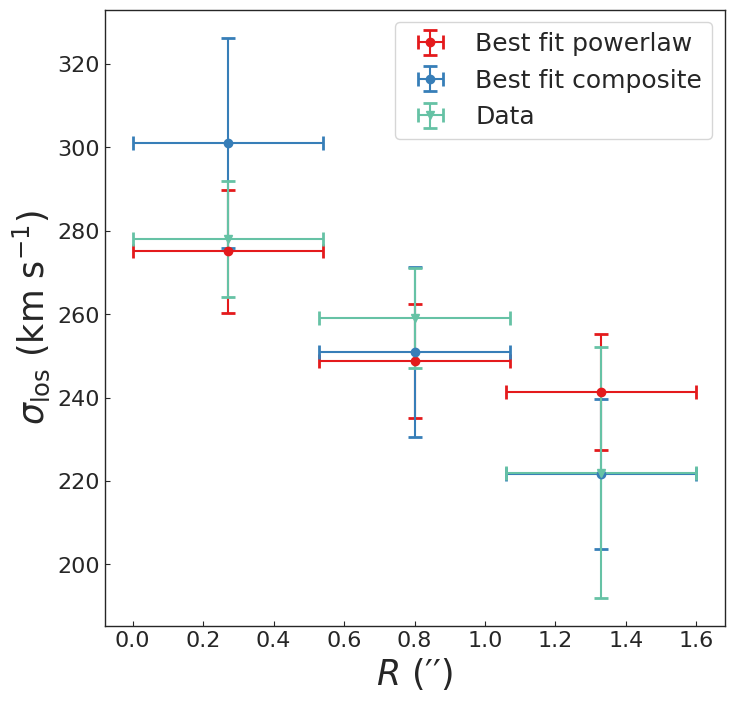}
    \endminipage
    \minipage{.3\textwidth}
    \includegraphics[width=\linewidth]{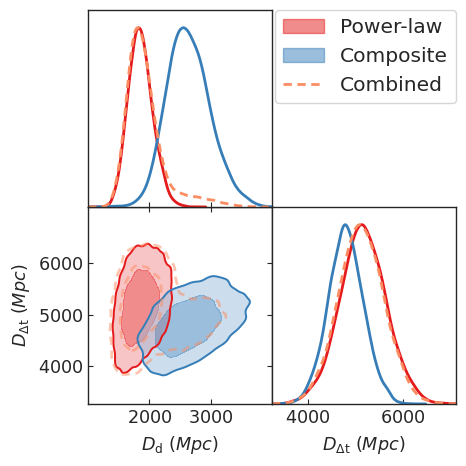}
    \endminipage
    \caption{ 
    \leftpanel Kinematic fit using power-law and composite models. Red and blue central dots show the best-fit values, while error bars are the 16th and 84th percentiles of 100 predicted velocity dispersions.   
    \rightpanel Posterior of \dd\ and  \ddt \ using kinematics based on the power-law and composite mass models. Using likelihood weighting we measured $\ddt\ =5083^{+504}_{487}$ Mpc and $\dd\ = 1921^{+254}_{-236}$ Mpc. }
    \label{fig:postkin}
    
\end{figure*}

Following the methodology of \citet{shajib23}, we importance sampled the posterior distribution of the lens model obtained with Eq.\ref{eq:weight}, $\xi_{\rm light},\xi_{\rm mass}$ together with the cosmological distances \dd\ and \ddt \ given the three radial bins of the velocity dispersion \slos and the external convergence \kext \ measured 
\begin{align}
    \mathcal{P}(\xi_{\rm light},\xi_{\rm mass}, \dd,\ddt \vert \slos, \kext) \nonumber \\    
    \propto\mathcal{L}(\slos \vert \slos_{\rm modelled}) p( \kext)
    \label{eq:finalposterior}
\end{align}
with p(\kext) the prior on \kext \ obtained in Section \ref{sec:kextmeas}.

The likelihood of the modelled 3-bin velocity dispersion is computed with the following 
\begin{align}
    & \mathcal{L}(\slos \vert \slos_{\rm modelled} ) \nonumber \\ 
    &  \propto  \exp\left[ -\frac{1}{2}(\slos-\slos_{\rm modelled})^{\rm T} ~ \mathcal{C}^{-1} ~ (\slos-\slos_{\rm modelled})\right],
\end{align}

where $\slos_{modelled}$ is the 3-bins velocity-dispersion computed with the {\tt JamPy} package\footnote{We used version 7.2 from \url{https://pypi.org/project/jampy/}} \citep{Cappellari2008, Cappellari2020} through Eqs. \ref{eq:slos} and \ref{eq:slosfinal}. $\mathcal{C}$ is the covariance matrix between \slos \ measured in each bin determined in Section \ref{sec:vdispmeas}. 
In practice, the software uses multi-gaussian-expansions \citep[MGE]{emsellem94,Cappellari2002} fits of the mass and light profiles to be able to deproject these surface densities into 3D ones in a straightforward way \citep{monnet92}.
The light MGE fit is based on the lens light profile corresponding to the $\xi_{\rm light}$ with maximum likelihood. The PSF is modelled as a Moffat with $\beta = 1.96$ and ${\rm FWHM} = 0.54 \arcs$ based on the mean value of these parameters along the wavelength range used for the \slos \ measurement  (i.e. [6000-9000] \AA) that is shown in the bottom panel of Fig.~\ref{fig:musecube}. 

As explained in Section \ref{sec:kinmod}, we adopted a uniform prior $\mathcal{U}(-0.12,0.13)$ on \bani \. The resulting kinematic fit for each modelling family is shown on the left panel of Fig.~\ref{fig:postkin} while the associated posterior distribution of \dd\ and \ddt\ is shown in the right panel. 
The kinematic modelling performed without allowing for the internal mass-sheet shows that the steepness of the composite model fails to correctly fit the observed velocity dispersion profile, especially in the innermost bin, where the best fit predicts 301 \kms, i.e 1.6 $\sigma$ away from the observed value. This results in a 2.1 $\sigma$ tension between the inferred \dd. The likelihood-weighted combination assigns 95 \% of the final posterior to the power-law model, effectively discarding the composite model measurement.

\begin{figure}[h!]
    \centering
    \includegraphics[width=.95\linewidth]{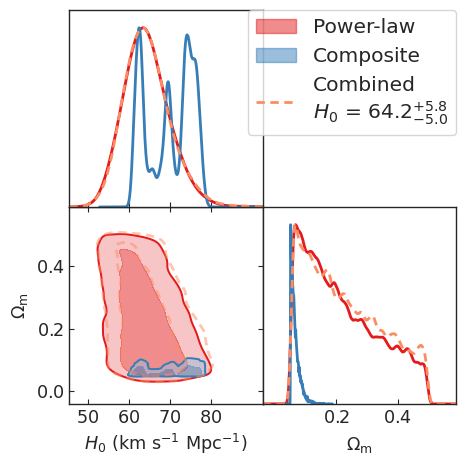}
    \caption{\hc \ measurement with \HEonze using the joint constraint of \ddt\ and \dd\ measured with the power-law, composite models, and both models combined. Our final measurement using the likelihood weighted combination of both models is $\hc = 64.2^{+5.8}_{-5.0} \kmsmpc$ for $\lint=1$.}
    \label{fig:h01104}
\end{figure}

Using the posterior distribution of \dd \ and \ddt \ we then inferred \hc \ in a flat-\lcdm \ cosmology with the purposedly conservative uniform priors $\mathcal{U}(0,150) \ \kmsmpc$ on \hc and $\mathcal{U}(0.05,0.5)$ on $\Omega_{\rm m}$. The resulting posterior distributions obtained for both models and the combination are displayed in Fig.~\ref{fig:h01104}. 
As with every TDCOSMO \hc measurement, the entirety of the analysis was performed blindly to avoid confirmation bias. The final value was revealed on July 8th of 2025, $\hc = 64.2^{+5.8}_{-5.0} $ \kmsmpc, i.e. an 8.5\% precision from a single doubly imaged lens.

\section{Discussion and conclusion \label{sec:discussion}}
Table~\ref{tab:he1104summary} summarises all the cosmography-related measurements achieved for \HEonze. 

\begin{table}
    \centering
    \caption{Summary of all the relevant \HEonze measurements.}
    \label{tab:he1104summary}
    \begin{tabular}{c|c|c|c}
    \multicolumn{4}{c}{Lens modelling measurements}\\[0.5ex]
    \hline
    \hline
         & Power-law & Composite & \thead{Relative \\ difference} \\
         \hline
         
        \thetae & $1.359 \pm 0.007$ & $1.369 \pm 0.012$ & 0.7 \%\\[0.4ex]
$\gamma$ & $1.85\pm 0.04$& $2.05 \pm 0.10$& 10.8 \%\\[0.4ex]
\gext & $0.14\pm0.01$& $0.12\pm 0.02$& 14.2\%\\[0.4ex]
\hline
\multicolumn{4}{c}{After combination with stellar kinematics}\\[0.5ex]
    \hline
    \hline
     & & & Combination\\[0.5ex]
     \hline
    $\ddt$ [Mpc] & $5131_{-494}^{+509}$ & $4783_{-295}^{+286}$ & $5083^{+504}_{487}$\\[0.5ex]
    \dd\ \ [Mpc] & $1854_{-195}^{+193}$ & $2559_{-269}^{+265}$ & $1921^{+254}_{-236}$\\[0.5ex]
\hline
    \multicolumn{4}{c}{Ancillary measurements}\\
    \hline
    \hline
    \dtab & \multicolumn{3}{c}{$176.3^{+11.4}_{-10.3}$ days}\\ [0.5ex]
    \kext & \multicolumn{3}{c}{$-0.019_{-0.020}^{+0.058}$}\\[0.5ex]
    $\slos_1$
     & \multicolumn{3}{c}{278 $\pm$ 7 (sys) $\pm$ 13 (stat) \kms} \\[0.4ex]
    $\slos_2$ & \multicolumn{3}{c}{259 $\pm$ 9 (sys) $\pm$ 8 (stat) \kms}\\[0.4ex]
    $\slos_3$ & \multicolumn{3}{c}{222 $\pm$ 8 (sys) $\pm$ 29 (stat) \kms }\\[0.4ex]

    \end{tabular}
    \tablefoot{The relative difference was computed using the power-law model. The combination of both models is a kinematic-fitting likelihood-weighted approach.}
    
\end{table}

\subsection{Time-delay}
The measurement of the time delay is marginally higher than previous estimates \citep[e.g.][measured \dtab = 162$\pm$6.1 days]{poindexter07, morgan08}. It relies on the longest and best-sampled light curve of that system, with the two most recent (ECAM and WFI) datasets having at least twice the cadence as the previous SMARTS dataset. It was estimated with a data-driven algorithm whose accuracy was assessed on multiple simulated and observed light curves, which is crucial, as we saw that the extrinsic variability is particularly challenging to model in this system. We therefore have the most robust estimate for this object. 

\subsection{Resolved kinematics}
The main limitation of the resolved velocity dispersion measurement comes from the proximity and brightness of image A to the lens galaxy. Nevertheless, by subtracting the quasar images, we were able to extract spectra from three concentric bins with S/N > 14 $\AA^{-1}$. Using state-of-the-art methodology to constrain the effect of stellar template fitting hyperparameters (polynomial degrees and stellar template libraries), we provided the velocity dispersion measurement in each bin. We ensured the robustness of these measurements against different spectral and spatial masks and proved that the lens is a slow rotator (see Appendix \ref{app:slowrotator}). The latter justifies the use of spherically aligned JAM for the kinematic modelling.

\subsection{Image and kinematic modelling}
The image reconstruction was performed with the two main parameterisation families: power-law and composite. Three configurations of nearby perturbers were accounted for individually, with the addition of SIS and NFW components, and the resulting measurements were marginalised upon these configurations.

The main difference between the two models resides in the slope of the radial mass profile $\gamma$. Such a difference was already observed in modelling the quadruply lensed quasar WGD 2038$-$4008 \citep{shajib22}, where the system's compactness only allowed to probe the inner regions of the mass profile.  
In the present case, the resolved velocity dispersion heavily favours the power-law model over the composite for this system. This may be because \HEonze's configuration offers too few constraints for a composite model, thus leading to potential biases or simply that the power law model is a better description over the range probed.
The discrepancy between the external convergence, \kext, and the external shear, \gext, inferred independently might be counter-intuitive in light of simulations' properties \citep[e.g.][]{tang25}. However, in this work \gext is primarily encompassing the effect of perturbers within 5\arcs of the main deflector, whereas \kext is inferred from galaxies up to 120\arcs away, excluding the ones within 5\arcs. As discussed in Section \ref{sec:lensmodel_posteriorsampling}, the large value of \gext may arise from genuine external shear or simply be inflated by inherent lens-modelling degeneracies. We could not disentangle the contribution of each effect with the available data, nor rule out unmodelled convergence. However, since the brightest perturbers we considered are already low-mass (see upper part of Table \ref{tab:perturbergroups}), any additional convergence from fainter perturbers should be small.

\subsection{Implications for time delay cosmography}

This work shows that the high precision of the time-delay measurement and the use of resolved kinematics allowed us to reach a precision of 8.5\% on \hc with a double lens quasar, which is comparable to previous TDCOSMO single lens measurements ranging from 4\% to 28\% \citep{millon20c, wong24}. 
The main limitation of this work is the fact that we fixed the internal mass-sheet degeneracy parameter, \lint = 1. This parameter will be constrained during the next joint hierarchical modelling of the TDCOSMO lens sample. The \hc value and precision of this system are therefore bound to change as \hc $\rightarrow$ \hc \lint, ($\delta$\hc / \hc)$^2 \rightarrow (\delta$\hc / \hc$)^2+\delta \lint ^2$.

Overall, this work is a stepping stone towards building a sample of double-lens quasar-based \hc measurement, which is vital to investigate potential biases of the TDCOSMO method and to accelerate progress towards a 1\% measurement of \hc. 

\label{sec:conclusion}

\begin{acknowledgements}

E.P. is supported by JSPS KAKENHI Grant Number JP24H00221.
F.C. is supported by the Swiss National Science Foundation (SNSF) and by the European Research
Council (ERC) under the European Union’s Horizon 2020 research and innovation programme
(COSMICLENS: grant agreement No 787886). C.D.F. acknowledges support for this work from the National Science Foundation under Grant No. AST-2407278. A.G. acknowledges support by the SNSF (Swiss National Science Foundation) through a mobility grant P500PT\_211034. M.M. acknowledges support by the SNSF (Swiss National Science Foundation) through mobility grant P500PT\_203114 and return CH grant P5R5PT\_225598. D.S. acknowledges the support of the Fonds de la Recherche Scientifique-FNRS, Belgium, under grant No. 4.4503.1. S.B. acknowledges support by the Department of Physics and Astronomy, Stony Brook University. C.L. acknowledges funding from the European Union’s Horizon Europe research and innovation programme under the Marie Sklodovska-Curie grant agreement No. 101105725.
 A.J.S. received support from NASA through STScI grants HST-GO-16773 and JWST-GO-2974. S.H.S. thanks the Max Planck Society for support through the Max Planck Fellowship. This work is supported in part by the Deutsche Forschungsgemeinschaft (DFG, German Research Foundation) under Germany's Excellence Strategy -- EXC-2094 -- 390783311. T.T. acknowledges support by NSF through grants NSF-AST-1906976, NSF-AST-1836016 and NSF-AST-2407277, and from the Moore Foundation through grant 8548. K.C.W. acknowledges support by JSPS KAKENHI Grant Numbers JP24K07089, JP24H00221. V.M. acknowledges support from ANID FONDECYT Regular grant number 1231418, Centro de Astrof\'{\i}sica de Valpara\'{\i}so CIDI 21, and ANID, Millennium Science Initiative, AIM23-0001.
 This work used observations collected at the European Southern Observatory under ESO programmes 092.0515(B) and 0102.A-0600(A). 
This work used Python packages {\tt Numpy} \citep{numpy}, {\tt Astropy} \citep{astropy} , and  {\tt Scipy} \citep{scipy}
\end{acknowledgements}

\bibliographystyle{aa}
\bibliography{bib}

\begin{appendix}

\section{Velocity dispersion measurement additional material}
\subsection{PSF subtraction in the MUSE cube \label{appendix:musepsfsub}}

\begin{figure}[h!]
\centering
    \includegraphics[width=\linewidth]{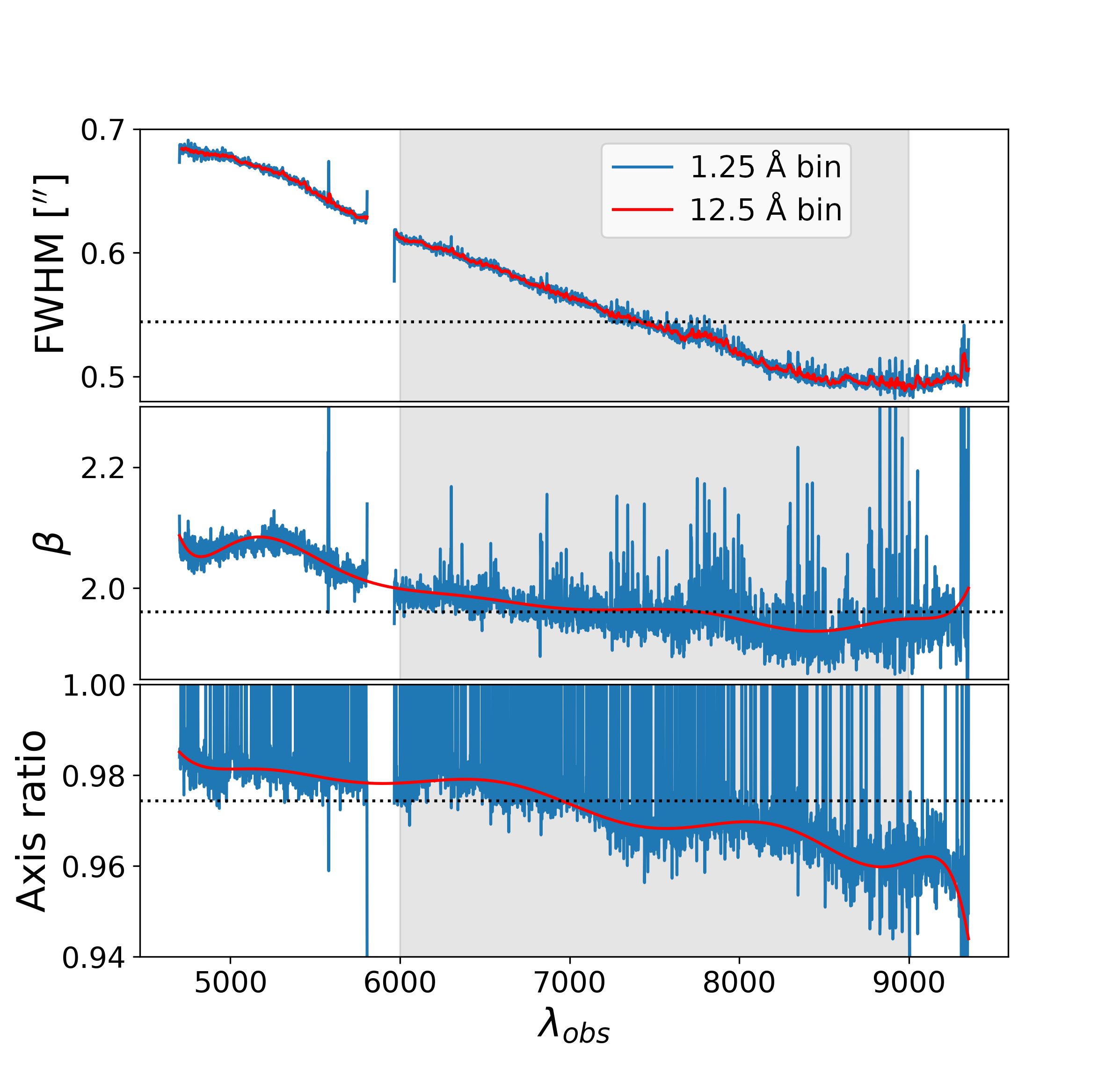}
    \caption{PSF parameters fitted across wavelength. The values obtained when using stacks of 12.5 \AA are used as constraining prior for the wavelength-by-wavelength fit. The shaded area indicates the range of wavelength used for the \slos \ measurement, and the dotted lines highlight the mean value of the parameters in this range.\label{fig:musecube_psfsub}}
    
\end{figure}

As shown in Fig.~\ref{fig:musecube_psfsub}, we first performed the fit on ten stacked wavelengths to constrain the parameters as a function of wavelength when performing the frame-by-frame fit. The wavelength change of the Moffat index $\beta$ and the Full-Width-Half-Maximum (FWHM) are consistent with expected MUSE PSF modelling \citep[e.g.][]{fusco20, weilbacher20}.  The relative instability of the ellipticity of the PSF can be attributed to the abrupt changes in the quasar and lens light shape due to emission and absorption lines.

\subsection{Is the lens rotating? \label{app:slowrotator}}

\begin{figure}
    \centering
    \includegraphics[width=\linewidth]{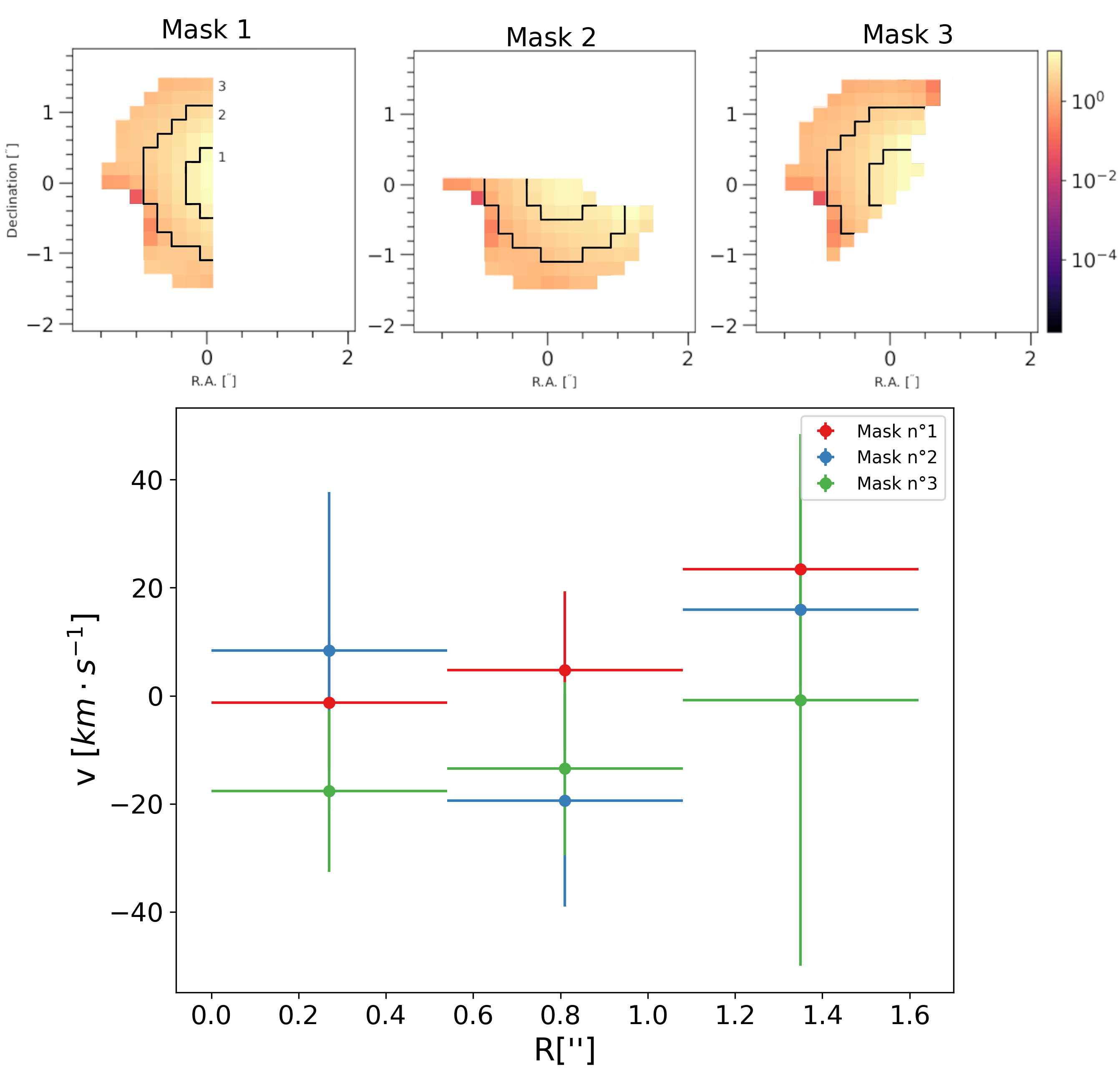}
    \caption{\toppanel three different masked MUSE cube cutouts. \bottompanel measured velocity of the three bins in each case. }
    \label{fig:vdipsmasks}
\end{figure}

To investigate bias in the \slos measurement due to possible rotation of the lens galaxy, we designed three spatial masks cutting half of the lens light area as shown in the left panel of \ref{fig:vdipsmasks}. The three masks are chosen to ensure an equivalent S/N for the three configurations while minimising the spatial overlap. With the same methodology as described in Section \ref{sec:vdispmeas}, we then measured the line-of-sight velocity of each bin $v^{\rm LOS}_{1,2,3}$.
The bottom panel of Fig.\ref{fig:vdipsmasks} shows that there is no sign of ordered rotation. These observations are, hence, consistent with the assumption that the lens galaxy is a slow rotator.

\section{MUSE by-products}
\subsection{Spectra extracted from the MUSE data \label{appendix:muse_spec}}

We present here the spectra of the center of the lens galaxy (S/N $\sim$ 30$\AA^{-1}$), perturbers P5 ($\sim$ 11$\AA^{-1}$), P6 ($\sim$ 7$\AA^{-1}$), and images A ($\sim$ 529$\AA^{-1}$) and B ($\sim$ 266$\AA^{-1}$) of the quasar extracted at the positions displayed in Fig.\ref{fig:musecube}.

\begin{figure*}

    \begin{subfigure}{0.95\textwidth}
    \centering
        \includegraphics[width=1.\linewidth]{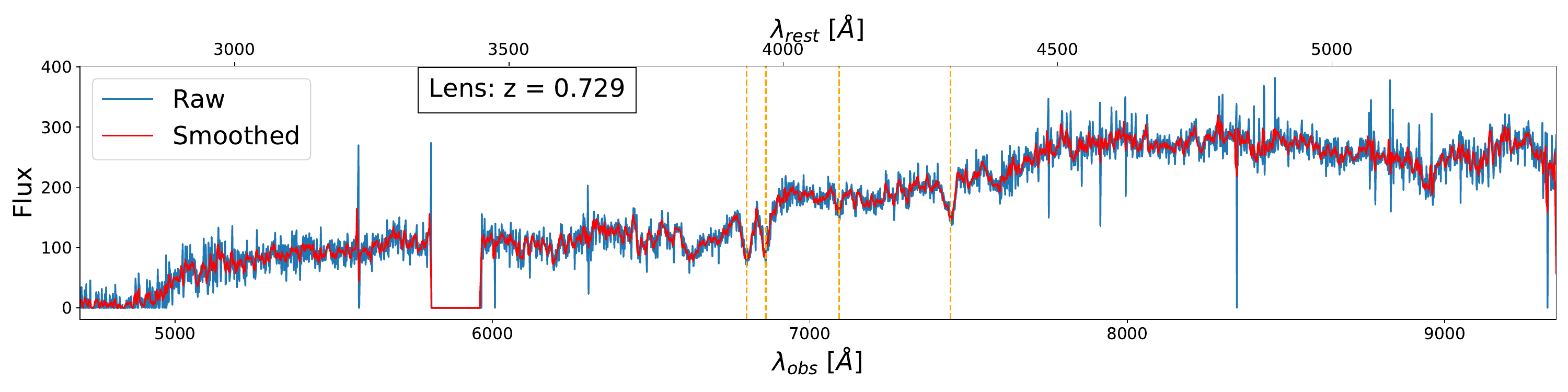}
    \end{subfigure}

    \begin{subfigure}{0.95\textwidth}
    \centering
        \includegraphics[width=\linewidth]{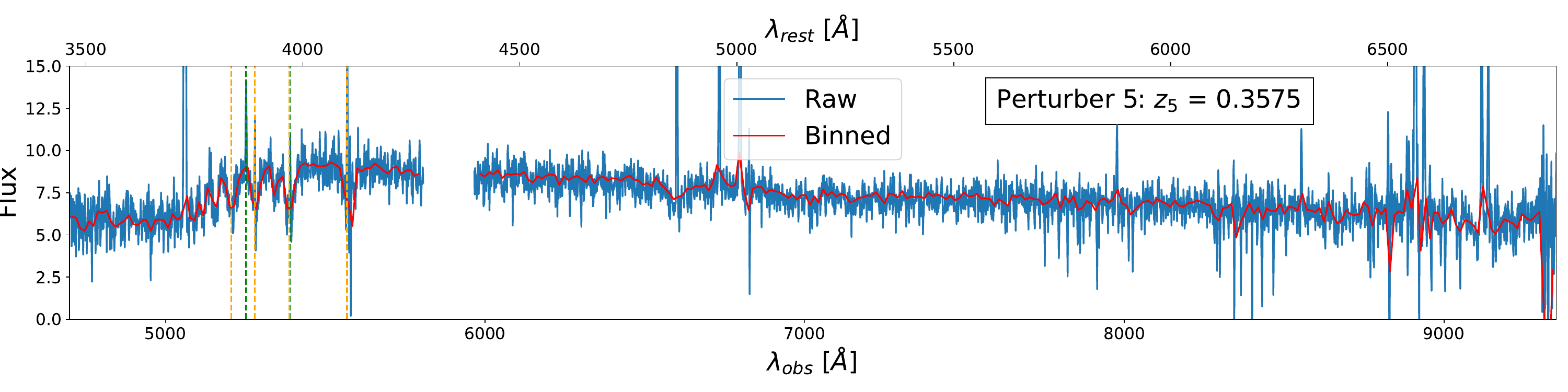}
    \end{subfigure}
    \begin{subfigure}{0.95\textwidth}
    \centering
        \includegraphics[width=\linewidth]{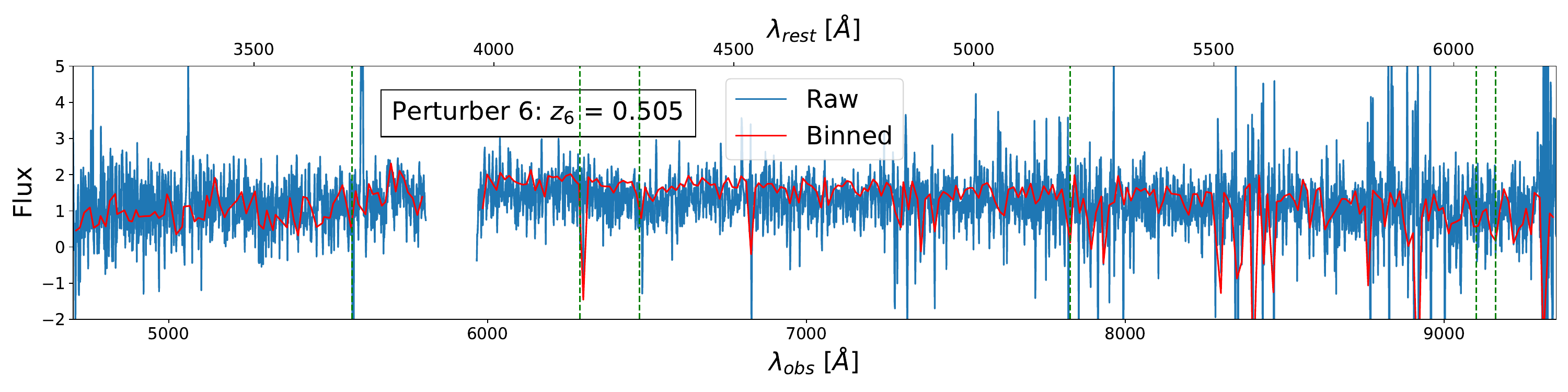}
    \end{subfigure}
    
    \begin{subfigure}{0.95\textwidth}
    \centering
        \includegraphics[width=1.05\linewidth]{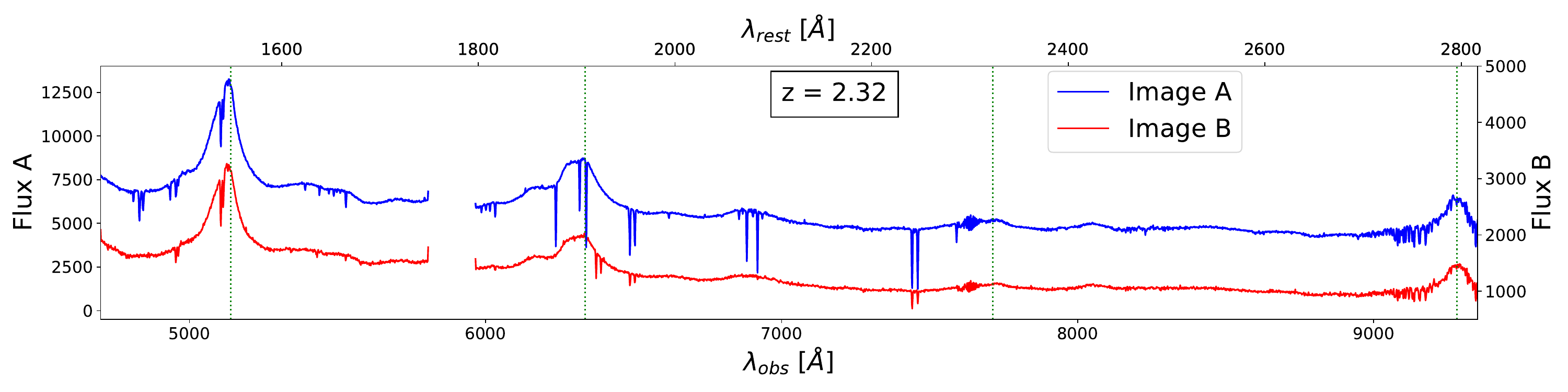}
    \end{subfigure}
    
    \caption{Integrated spectra from the MUSE data cube with apertures shown in Fig.~\ref{fig:musecube}. The binned flux is computed by imposing a minimal S/N of 5 per wavelength bin. From top to bottom, Lens centre: from left to right, the stellar absorption lines \ion{Ca}{H}, \ion{Ca}{K}, \ion{H}{$\delta$}, and G-band (orange dashed lines) are consistent with the previously determined $z_{\rm d}= 0.729$. 
    P5: from left to right, the galactic emission lines \ion{Ne}{iii}, \ion{He}{i}, \ion{H}{$\epsilon$}, and \ion{H}{$\delta$} are highlighted with green dashed lines and allowed us to measure $z_5=0.3575$. We also identify the stellar absorption lines \ion{H}{$\gamma$}, \ion{HE}{i}, \ion{Ca}{H}, and \ion{Ca}{K} allowing a velocity dispersion measurement. 
 P6: from left to right the lines \ion{Mg}{i}, \ion{Ca}{ii} \ion{O}{ii}, \ion{H}{$\beta$}, and \ion{O}{iii} doublet allowed us to measure $z_6=0.505$. Quasar images: from left to right, we recognise the \ion{C}{iv}, \ion{C}{iii}, \ion{C}{ii}, and \ion{Mg}{ii} emission lines. }
 \label{fig:spectra}
    
\end{figure*}

\subsection{Velocity dispersion of the perturber P5 \label{app:slosp5}}

\begin{figure*}[h!]
\centering
    \centering
    \includegraphics[width=\linewidth]{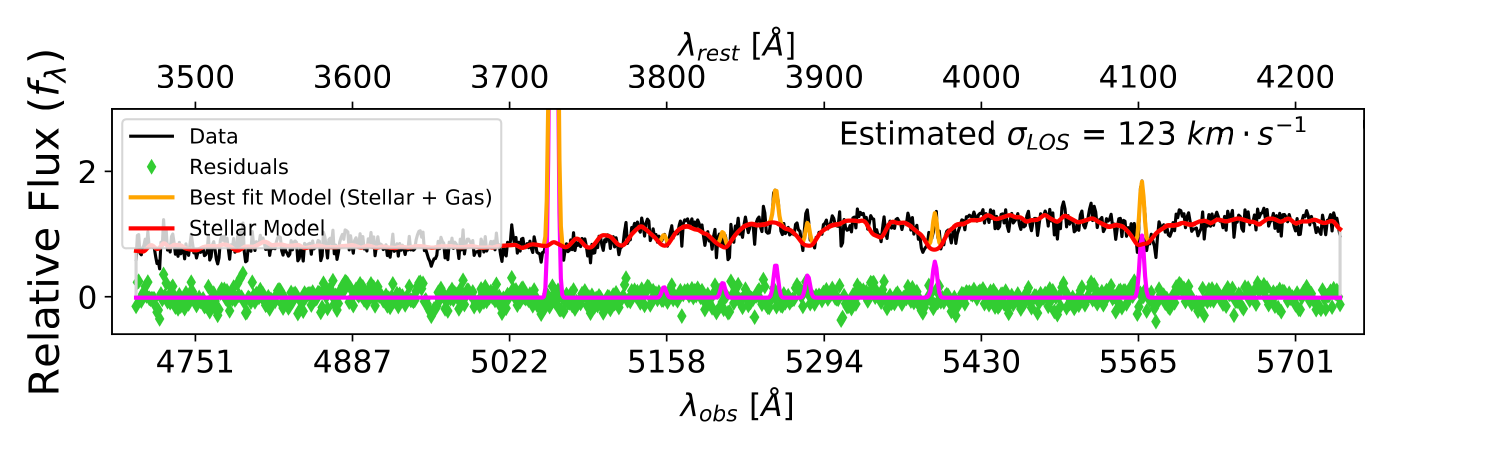}
    \caption{Measurement of the P5 stellar kinematics. Point-estimate of \slos$_{\rm P5}$. } \label{fig:vdisppert_pointestimate}
\end{figure*}

We applied the same stellar template fitting process as performed with the lens galaxy spectra to the spectrum of P5 shown in Fig.~\ref{fig:spectra}. In this case, we used the observed wavelength range [(4600,4700):(5600,5700)]$~\AA$. As shown by Fig.~\ref{fig:vdisppert_pointestimate}, the absorption lines are not altered by the sky but by galactic gas emission lines. We thus incorporated those in the model without testing different masks. We obtained the $\slos_{P5} = 123 \pm 71  \ \kms$.
According to \citet{knabel25}, spectra with relatively low S/N, such as this one, may be affected by systematic bias up to $\sim 4\%$, which is negligible compared to the uncertainty of the measurement.

\section{Lens modelling, additional material}
\subsection{PSF reconstruction for HST imaging}

\begin{figure}[h!]
     \centering
     \includegraphics[width = \linewidth]{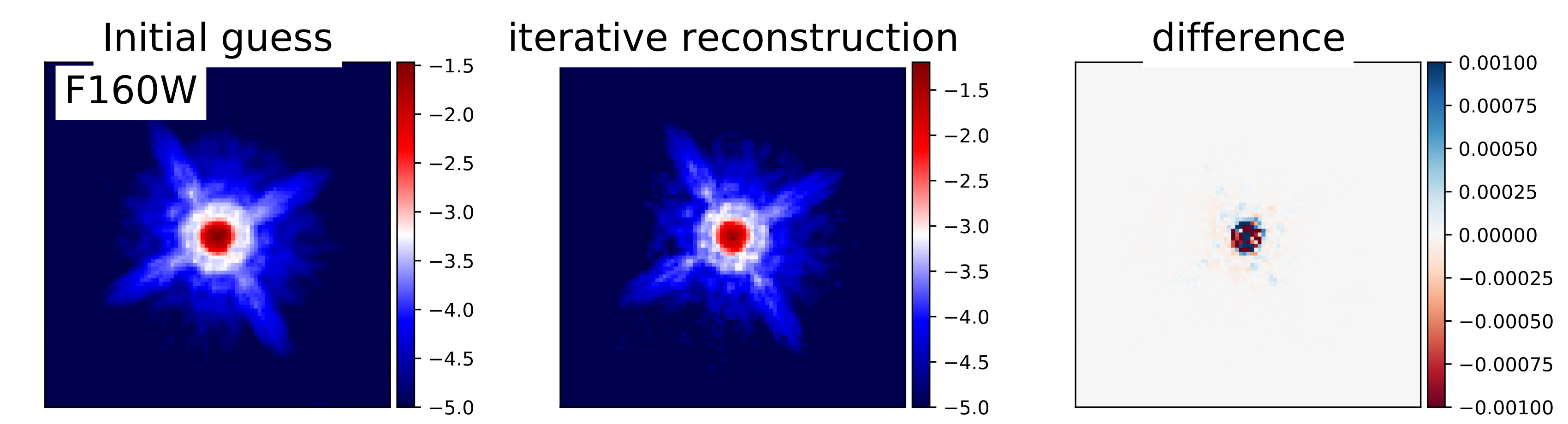}
     \caption{From left to right: Initial guess of the PSF from stacked stars in the field, final PSF obtained, and the difference between both. \label{fig:psf}}
 \end{figure}

The Fig.\ref{fig:psf} shows the PSF obtained after the iterative reconstruction procedure described in Section \ref{sec:psf}.

\subsection{Image reconstruction}

\begin{figure}
    \centering
    \includegraphics[width=\linewidth]{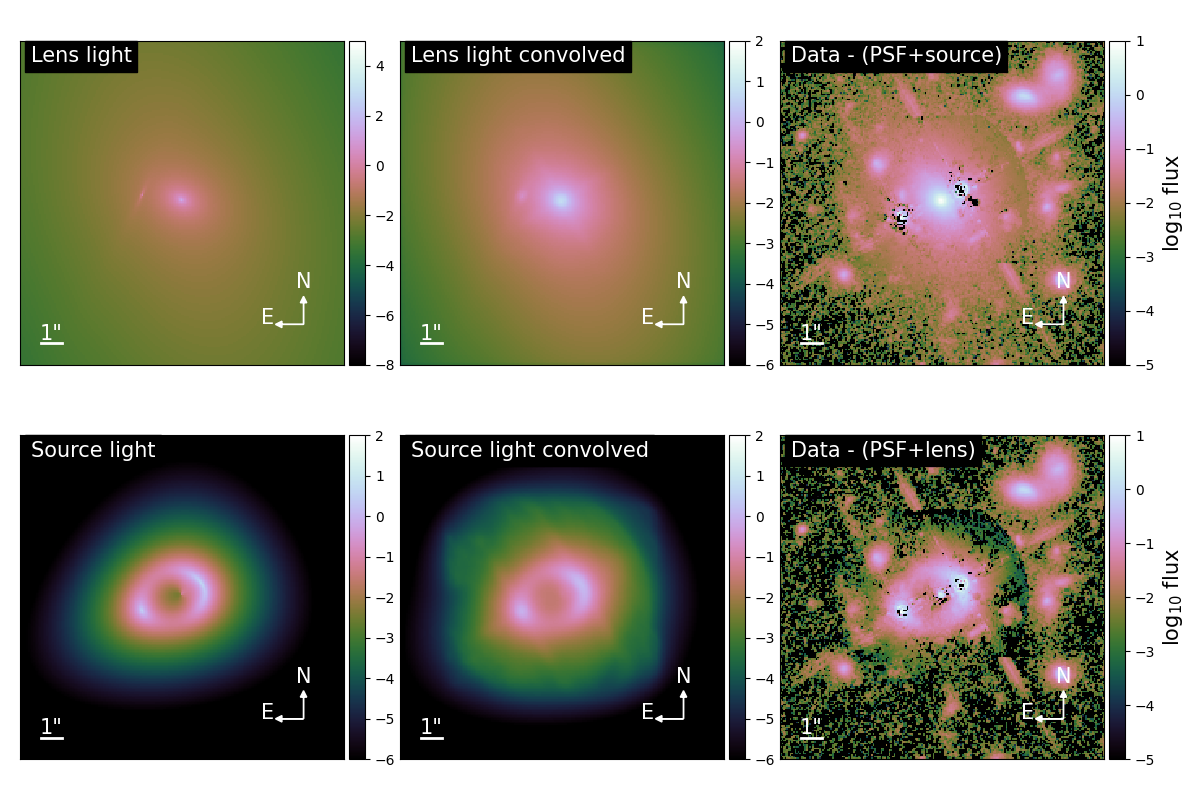}
    \caption{\HEonze reconstructed light component with power-law. From left to right: The light component unconvolved, convolved with the PSF, and its corresponding feature in the imaging data. The top row shows the lens light profile, and the bottom row shows the lensed source light profile. This figure shows that the light-components model corresponds to expectations based on the observation. The centre of the PSF and of the lens are not ideally modelled, which justifies the noise map increase in the first region and the masking of the second region.}
    \label{fig:decomppl}
\end{figure}

\begin{figure}
    \centering
    \includegraphics[width=0.9\linewidth]{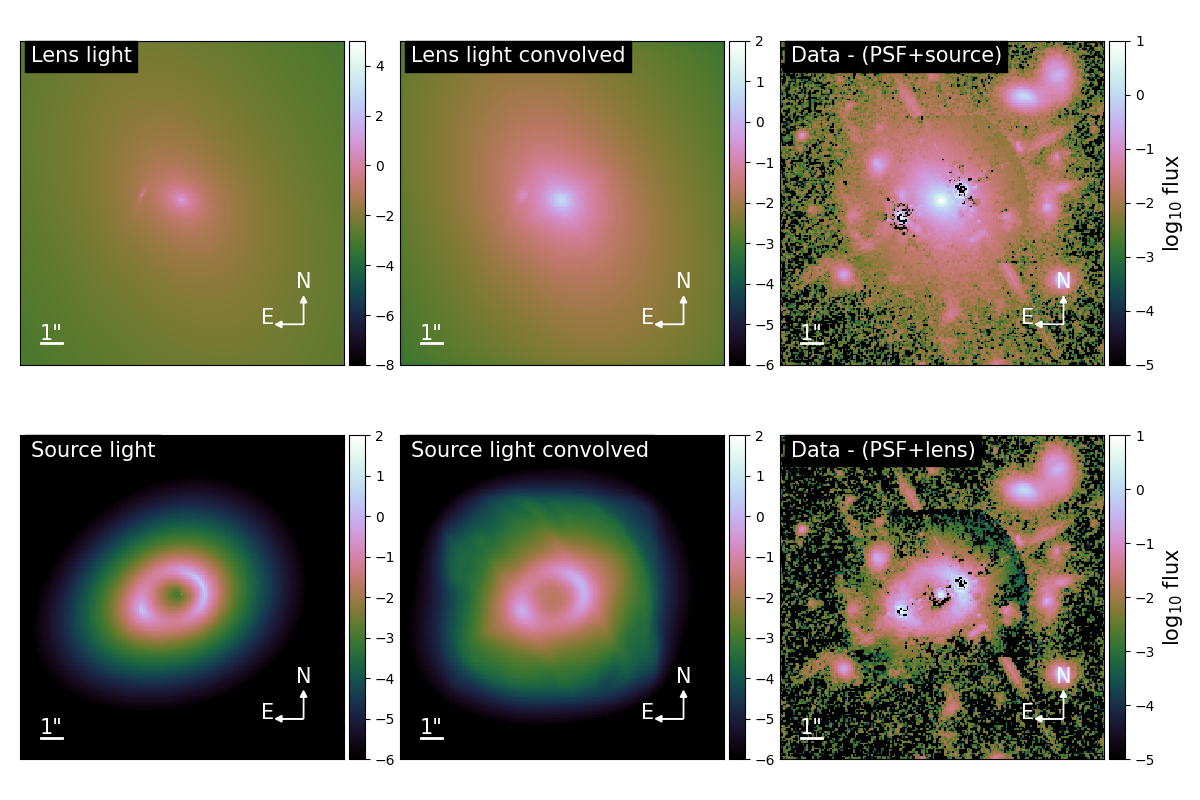}
    \caption{\HEonze reconstructed light component with power-law. From left to right: The light component unconvolved and then convolved with the PSF and its corresponding feature in the imaging data. The top row shows the lens light profile, and the bottom row shows the lensed source light profile. This figure shows that the light-components model corresponds to expectations based on the observation. The centre of the PSF and of the lens are not ideally modelled, which justifies the noise map increase in the first region and the masking of the second region.}
    \label{fig:decompcomp}
\end{figure}

We now show the decomposition of the imaging reconstruction, i.e. the lens light and source light obtained with a power-law in Fig.\ref{fig:decomppl} and composite in Fig.~\ref{fig:decompcomp} models. 

\section{Galaxy perturber groups members}
\begin{figure*}
	\centering
	
    \begin{subfigure}{0.3\textwidth}
    \label{fig:g1}
    \includegraphics[width=\textwidth]{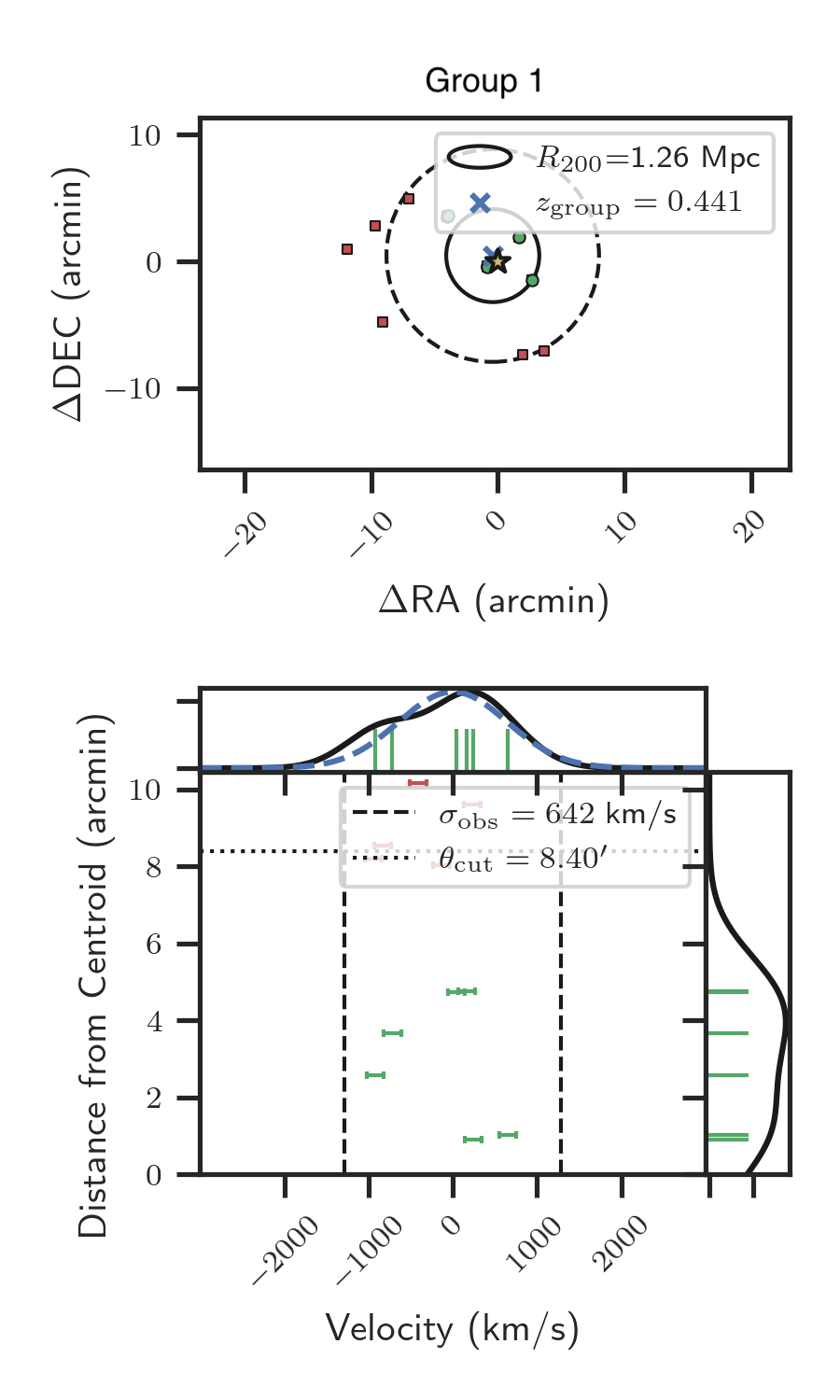}
    \end{subfigure}
    \begin{subfigure}{0.3\textwidth}
    \label{fig:g2}
    \includegraphics[width=\textwidth]{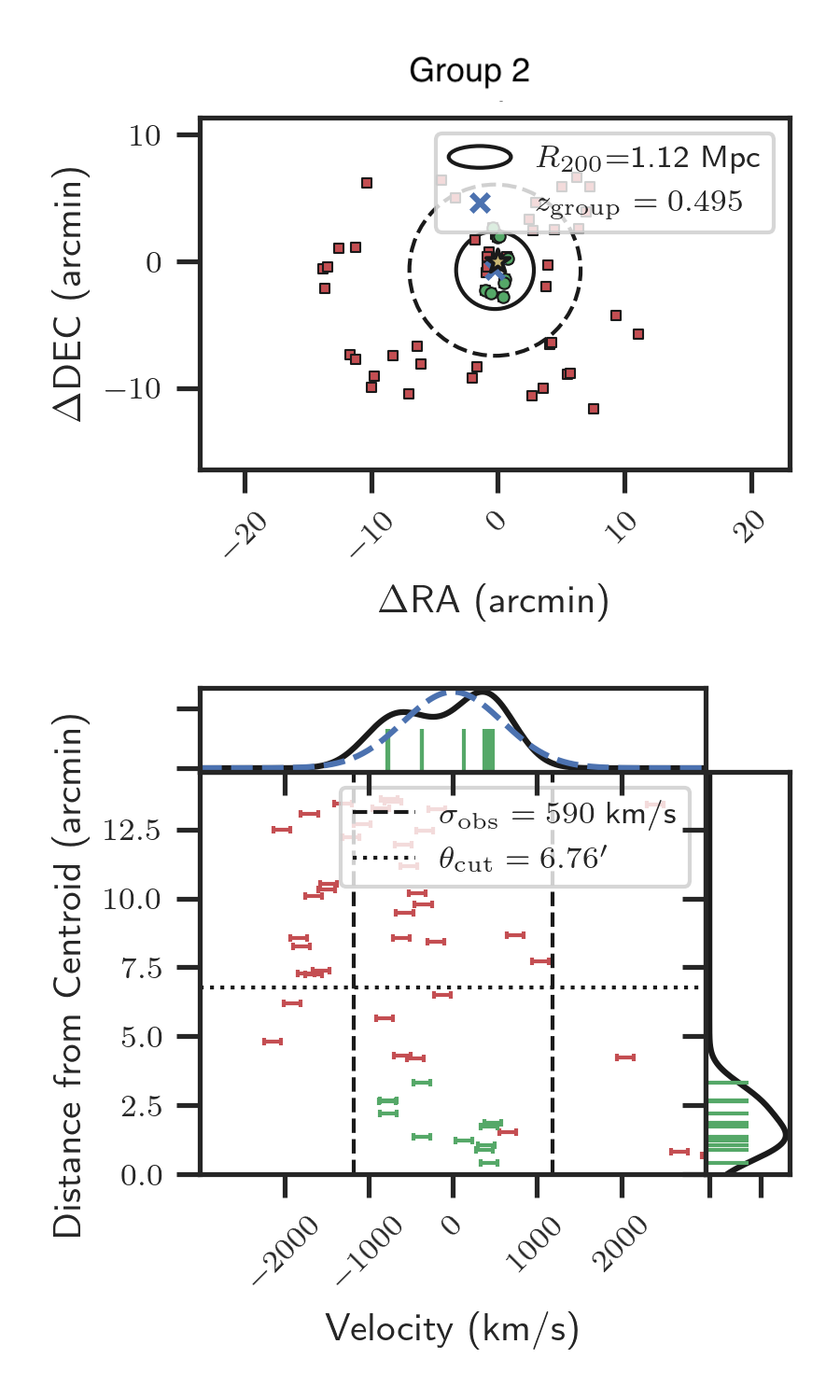}
    \end{subfigure} 
    \begin{subfigure}{0.3\textwidth}
    \label{fig:g3}
    \includegraphics[width=\textwidth]{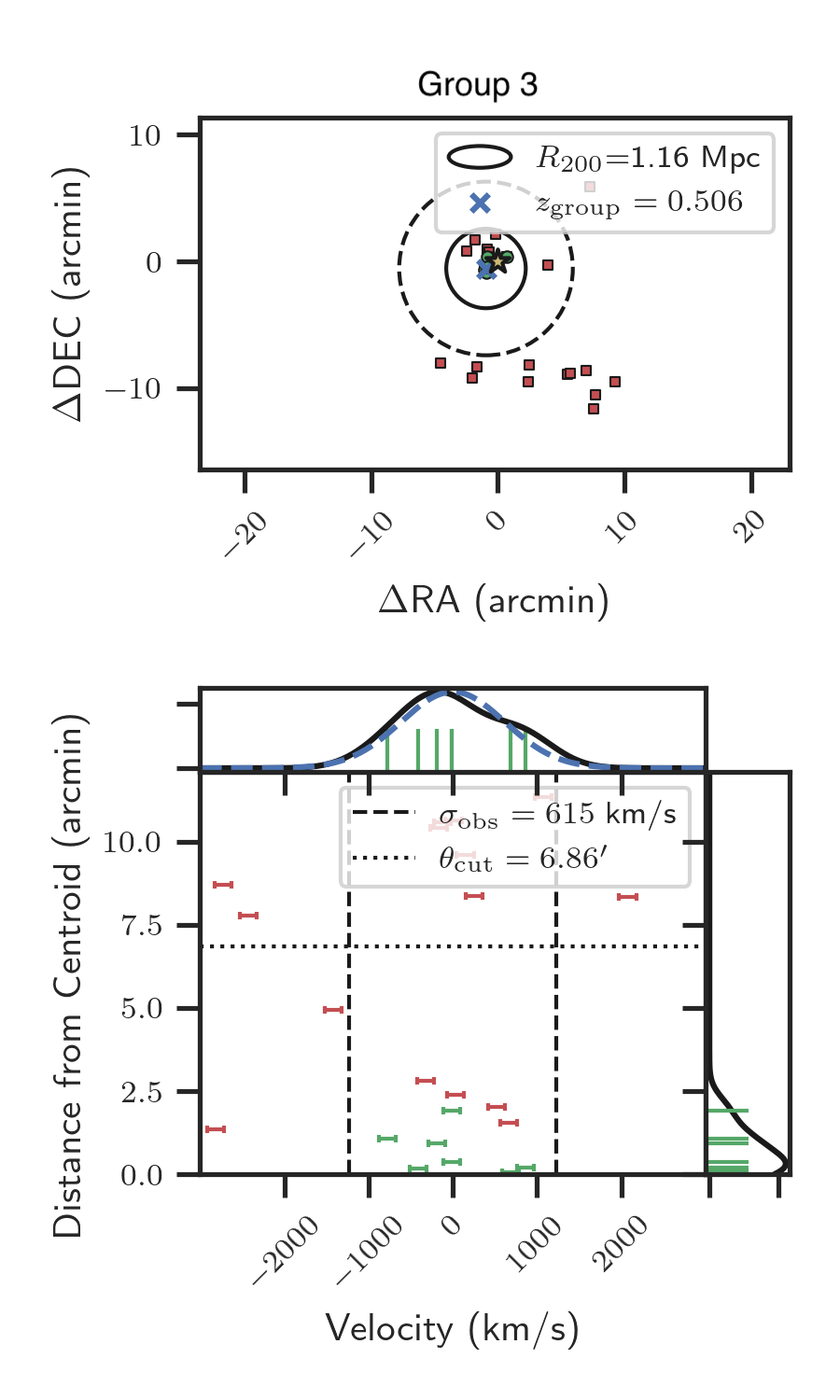}
    \end{subfigure}   
    \begin{subfigure}{0.3\textwidth}
    \label{fig:g4}
    \includegraphics[width=\textwidth]{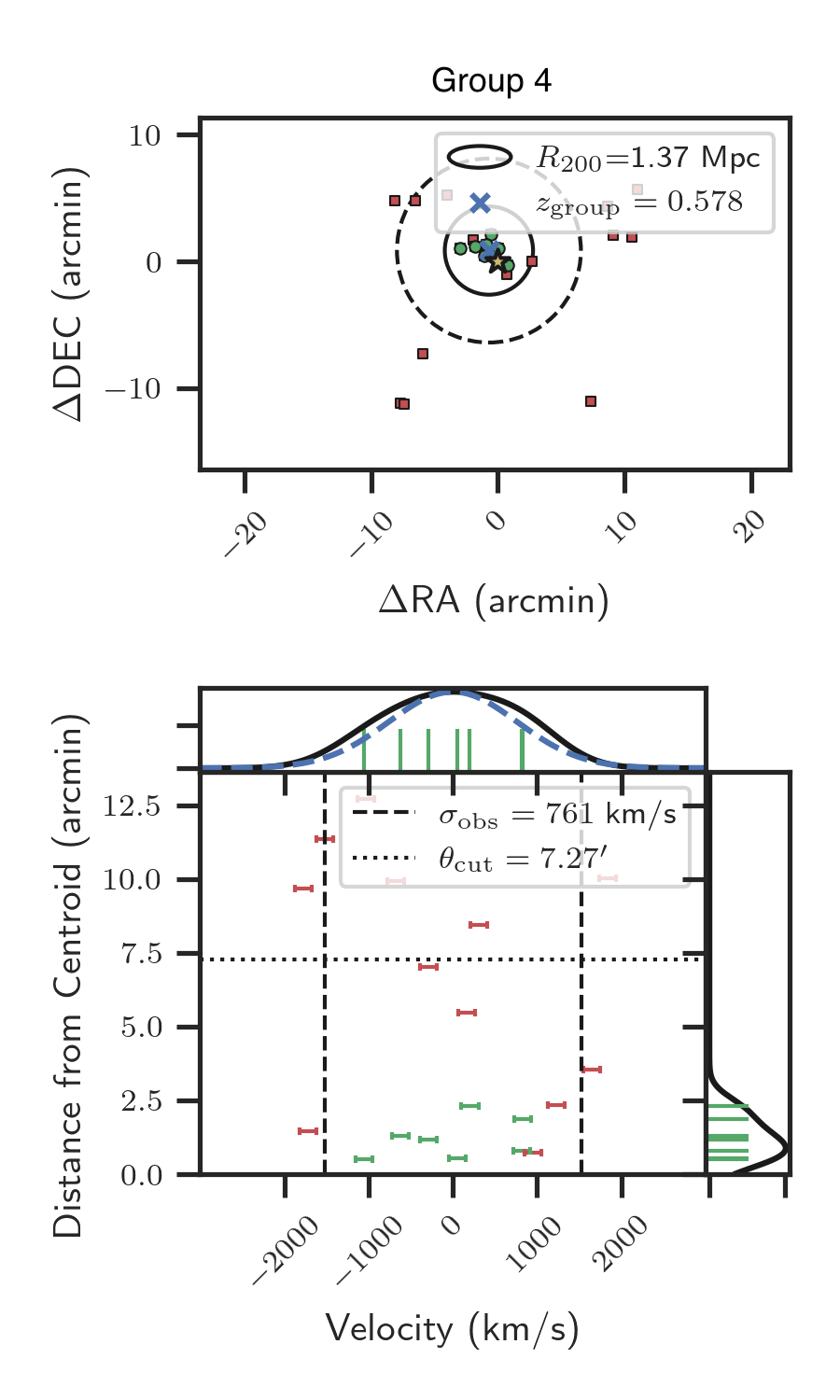}
    \end{subfigure}    
    \begin{subfigure}{0.3\textwidth}
    \label{fig:g5}
    \includegraphics[width=\textwidth]{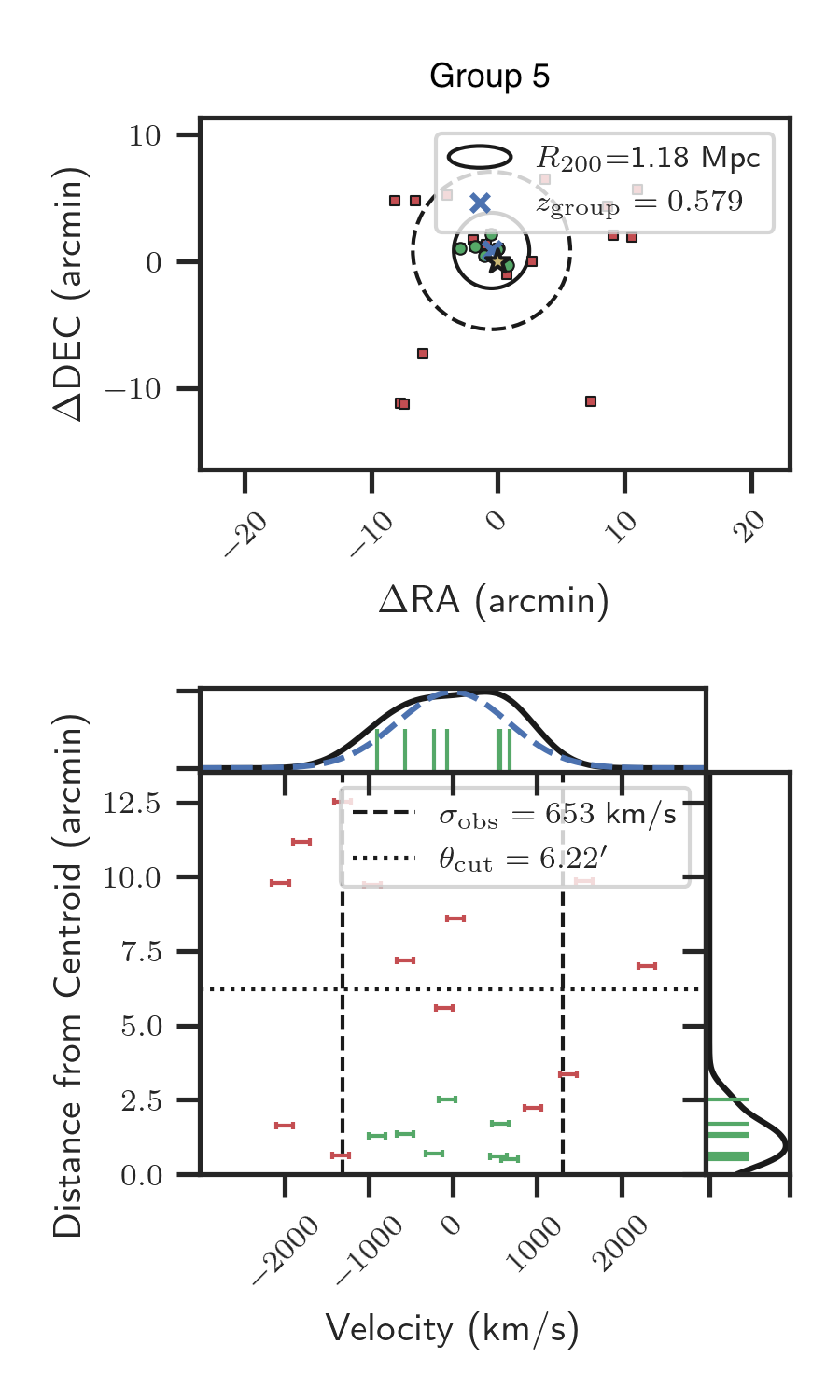}
    \end{subfigure}   
    \begin{subfigure}{0.3\textwidth}
    \label{fig:g6}
    \includegraphics[width=\textwidth]{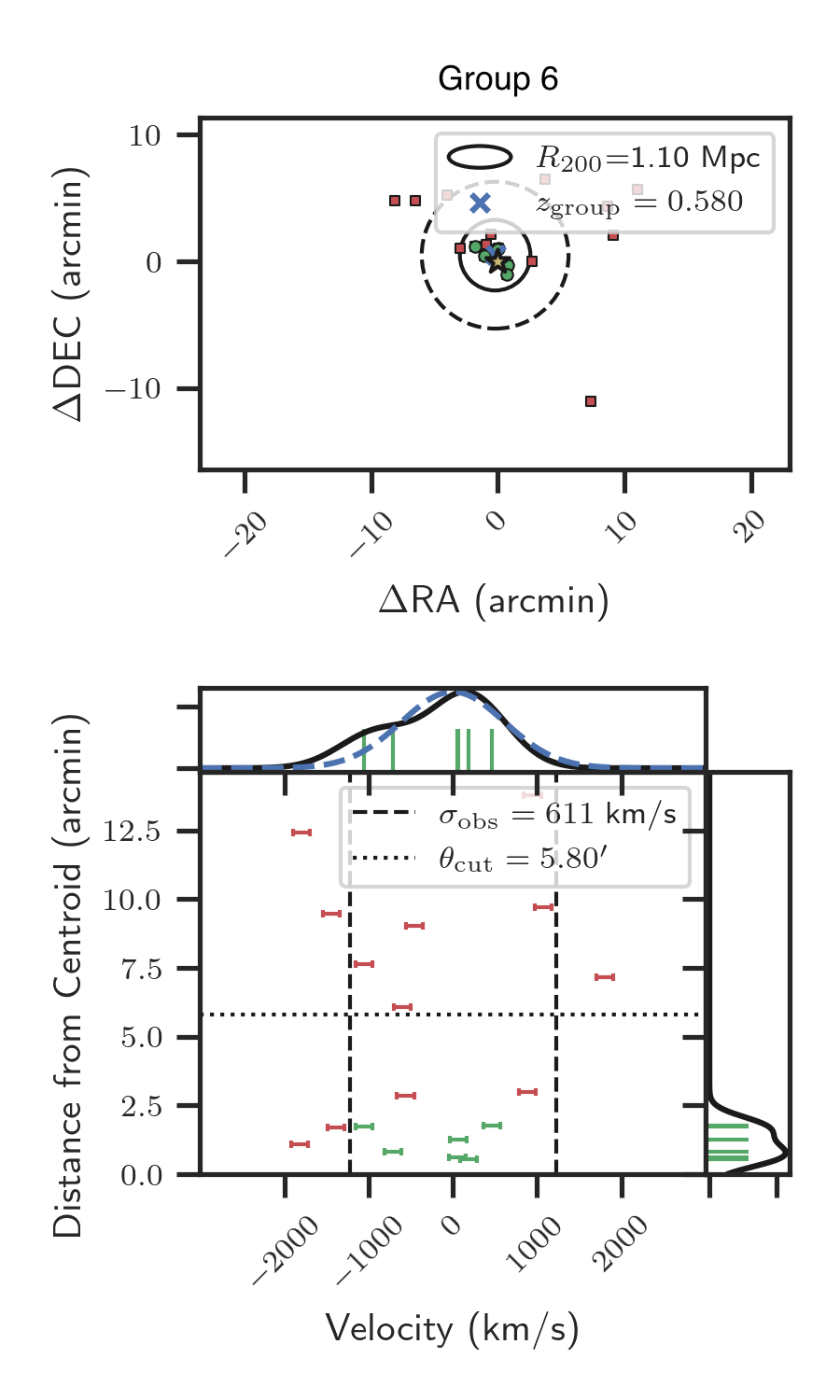}
    \end{subfigure}
    
 	\caption{ Galaxy groups identified in the spectroscopic sample of galaxies in the field of \HEonze. For each group, the upper plot shows the positions of the candidate member galaxies associated with that group relative to the lens galaxy, with rejected group members represented as red squares and accepted group members represented as green circles. The lens galaxy (star) and group centroid (cross) are also displayed. The $R_{200}$ radius of the group is represented by a solid line, while the dashed circle represents the angular separation cut of the group-finding algorithm in its final iteration. The lower plot shows the observer-frame velocity of individual member galaxies relative to the group centroid as a function of that galaxy's angular distance from the centroid. Galaxies that passed the iterative algorithm described in \citet{buckleygeer20}, \citet{sluse17}, and \citet{sluse19} are shown in green, while trial galaxy members that were cut through the algorithm are shown in red.  Horizontal error bars represent the measurement error for each galaxy. The final observer-frame velocity dispersion and angular separation cuts from the group-finding algorithm are presented as dashed and dotted lines, respectively. We also show 1-D histograms and rug plots of the velocity and distance distributions of the member galaxies. The 1-D histograms are produced using a kernel density estimate (KDE) with a bandwidth determined using Scott's Rule. In the 1D velocity histogram, the dashed blue line shows a Gaussian with a width equal to the observer-frame velocity dispersion of the group.}
    \label{fig:groups}
\end{figure*}

Fig. \ref{fig:groups} displays the position and velocities of the galaxy groups included as perturbers in the lens model. 

\end{appendix}
 
\end{document}